\documentclass[iop,apj,twocolappendix]{emulateapj}
\usepackage{amsmath,amssymb,amstext}

\usepackage[breaklinks,colorlinks,citecolor=blue,linkcolor=magenta]{hyperref} 

\usepackage[all]{hypcap} 
\usepackage{multirow}
\usepackage{booktabs}
\usepackage{enumitem}
\usepackage{rotating}

\usepackage{aas_macros}
\usepackage{natbib}
\usepackage[percent]{overpic}
\usepackage{lineno}
\usepackage[T1]{fontenc}

\shorttitle{Star Formation History of UGC~4483}

\shortauthors{Sacchi et al.}

\begin{document}

\title{Reaching the oldest stars beyond the Local Group: \\ ancient star formation in UGC~4483 \footnotemark[$\star$]} \footnotetext[$\star$]{Based on observations obtained with the NASA/ESA \textit{Hubble Space Telescope} at the Space Telescope Science Institute, which is operated by the Association of Universities for Research in Astronomy under NASA Contract NAS 5-26555.}
\author{Elena Sacchi$^{1,2,3}$,
Alessandra Aloisi$^{1}$,
Matteo Correnti$^{1}$,
Francesca Annibali$^{3}$,
Monica Tosi$^{3}$,
Alessia Garofalo$^{3}$,
Gisella Clementini$^{3}$,
Michele Cignoni$^{3,4,5}$,
Bethan James$^{1}$,
Marcella Marconi$^{6}$,
Tatiana Muraveva$^{3}$,
and Roeland van der Marel$^{1,7}$
}
\affil{
$^{1}$Space Telescope Science Institute, 3700 San Martin Drive, Baltimore, MD 21218, USA\\
$^{2}$ Leibniz-Institut für Astrophysik Potsdam, An der Sternwarte 16, 14482 Potsdam, Germany; esacchi@aip.de\\
$^{3}$INAF--Osservatorio di Astrofisica e Scienza dello Spazio di Bologna, Via Gobetti 93/3, I-40129 Bologna, Italy \\
$^{4}$Dipartimento di Fisica, Universit\`a di Pisa, Largo Bruno Pontecorvo, 3, 56127 Pisa, Italy \\
$^{5}$INFN, Sezione di Pisa, Largo Pontecorvo 3, 56127 Pisa, Italy \\
$^{6}$INAF--Osservatorio Astronomico di Capodimonte, Salita Moiariello 16, 80131, Naples, Italy \\
$^{7}$Center for Astrophysical Sciences, Department of Physics \& Astronomy, Johns Hopkins University, Baltimore, MD 21218, USA
}

\begin{abstract}
We present new WFC3/UVIS observations of UGC~4483, the closest example of a metal-poor blue compact dwarf galaxy, with a metallicity of $Z \simeq 1/15\ Z_{\odot}$ and located at a distance of $D \simeq 3.4$~Mpc. The extremely high quality of our new data allows us to clearly resolve the multiple stellar evolutionary phases populating the color-magnitude diagram (CMD), to reach more than 4 mag deeper than the tip of the red giant branch, and to detect for the first time core He-burning stars with masses $\lesssim 2$~M$_{\odot}$, populating the red clump and possibly the horizontal branch (HB) of the galaxy. By applying the synthetic CMD method to our observations, we determine an average star formation rate over the whole Hubble time of at least $(7.01\pm0.44) \times 10^{-4}$~$\mathrm{M_{\odot}/yr}$, corresponding to a total astrated stellar mass of $(9.60\pm0.61)\times 10^6$~$\mathrm{M_{\odot}}$, 87\% of which went into stars at epochs earlier than 1~Gyr ago. With our star formation history recovery method we find the best fit with a distance modulus of DM~=~$27.45\pm0.10$, slightly lower than previous estimates. Finally, we find strong evidence of an old ($\gtrsim 10$~Gyr) stellar population in UGC 4483 thanks to the detection of an HB phase and the identification of six candidate RR Lyrae variable stars.
\end{abstract}


\maketitle

\section{Introduction}
\setcounter{footnote}{16}
\urlstyle{sf}

Star formation (SF) studies beyond the Local Group have been pushed to their current limits in the past few years. Thanks to the spatial resolution and sensitivity of the \textit{Hubble Space Telescope} (\textit{HST}), it is possible to resolve single stars in galaxies up to $\sim 20$ Mpc, with the limitation of characterizing only the brightest stellar evolutionary features, i.e., the upper main sequence (MS), the He-burning phase of massive and intermediate-mass stars (blue loops, BLs), the asymptotic giant branch (AGB), and the red giant branch (RGB). 
Even though the RGB can host stars with any age between $1-2$  and 13 Gyr, it is unfortunately affected by an age-metallicity degeneracy; this implies that only the most recent star formation history (SFH), back to $1-2$~Gyr ago, can be derived with good time resolution ($\sim 10\%$), while a precise characterization of the SF behaviour is not possible at the oldest epochs.

\begin{figure*}
\centering
\includegraphics[width=0.49\linewidth]{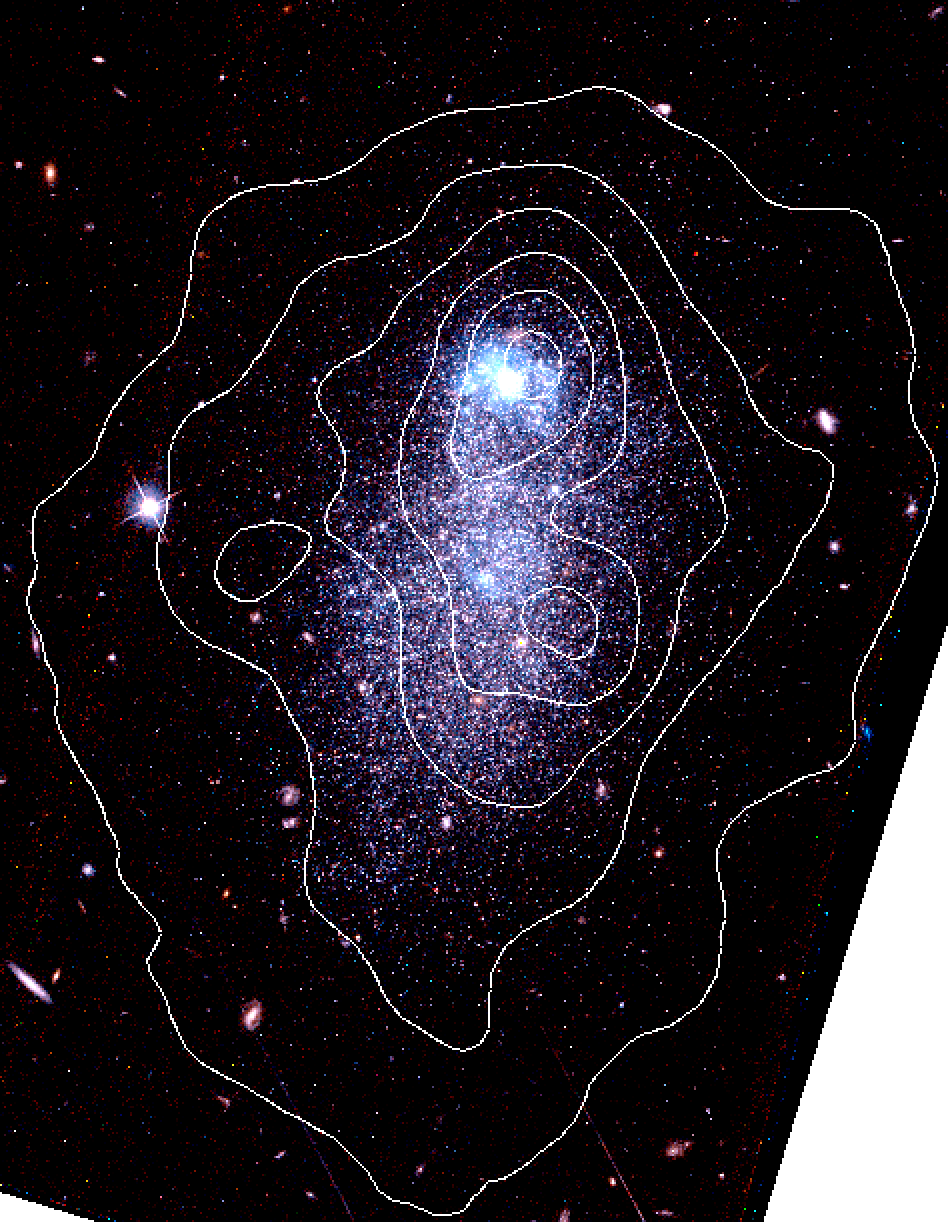}
\hfill
\includegraphics[width=0.49\linewidth]{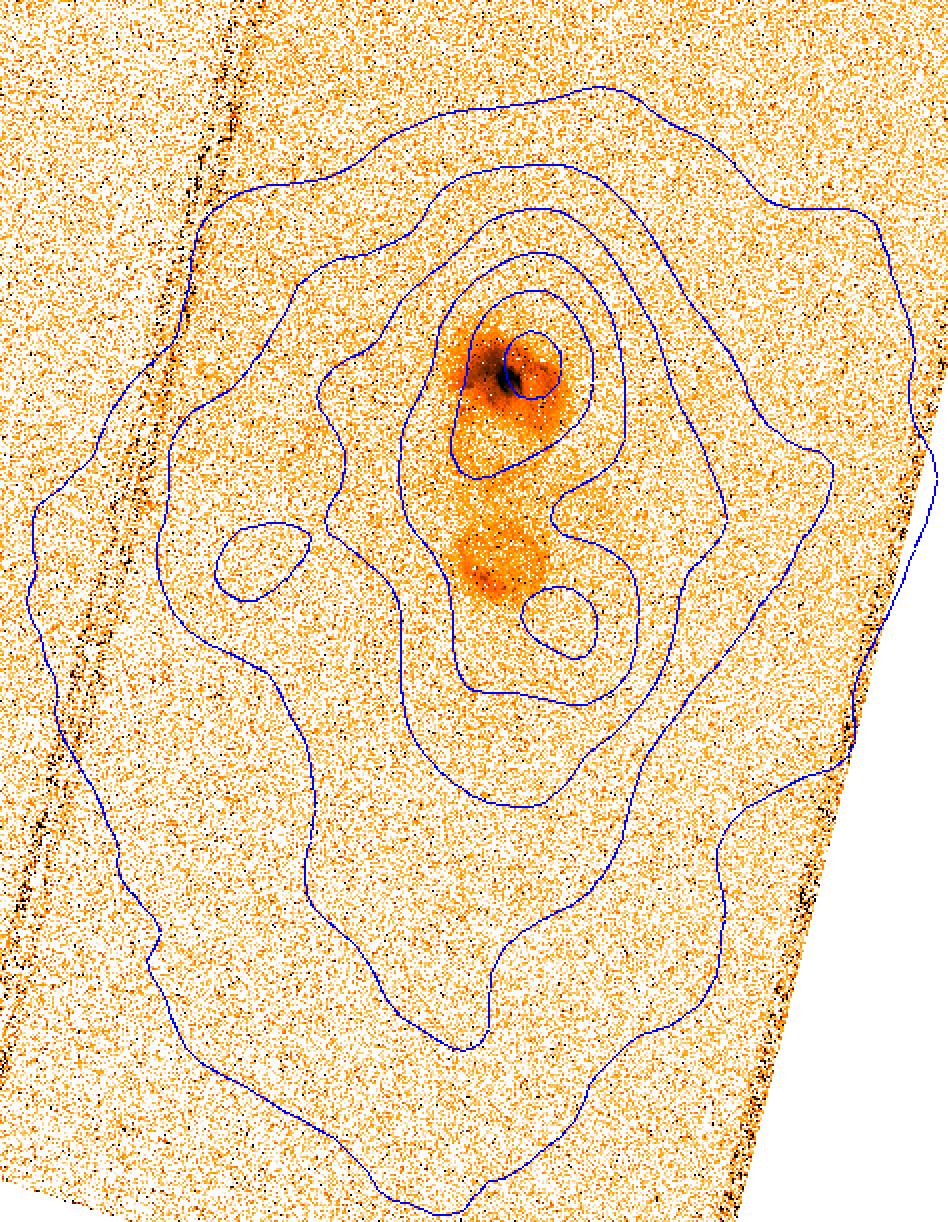}
\caption{\textit{Left panel.} 3-color composite image of UGC~4483 from our \textit{HST}/WFC3 observations: blue corresponds to F606W (broad $V$), red to F814W ($I$), while the green channel was obtained using the mean of the two. H\textsc{i} contours from the VLA-ANGST survey \citep{Ott2012} corresponding to column densities of $N_H=$ 3.0, 2.5, 2.0, 1.5, 1.0, and $0.5 \times 10^{21}$ cm$^{-2}$ have been superimposed to the HST image. The displayed field of view is $1.7' \times 2.2'$. North is up, East is left. \textit{Right panel.} F656N image of UGC~4483 with superimposed the same H\textsc{i} contours.}
\label{fig:image}
\end{figure*}

So, why do we even go through the trouble of studying such distant systems, with all the uncertainties associated with them? Beyond removing the environmental effects that the big spirals inside the Local Group have on smaller systems, and studying dwarf galaxies in isolation, a key reason is that a particular sub-group of the dwarf class is not present inside the Local Group: blue compact dwarf (BCD) galaxies. These are extremely interesting systems, most often characterized by bluer colors, more intense star formation activities, and higher central surface brightness compared to regular star-forming dwarfs. Because of their recent or ongoing bursts of SF and typical low metallicities, for a long time they were believed to be young galaxies, but all the systems studied in details through their color-magnitude diagrams (CMDs) and SFHs revealed a population of RGB stars, thus at least $1-2$ Gyr old, and possibly as old as 13.7 Gyr (see, e.g., \citealt{Schulte-Ladbeck2001,Annibali2003,Aloisi2007,McQuinn2015}).

All the studies conducted so far on BCDs are limited by the depth of the photometry, reaching $1-2$ magnitudes below the RGB tip (TRGB), which allows to robustly reconstruct the SFH up to a few Gyr ago only. To make progress, it is necessary to reach fainter features with discriminatory power at older ages. The brightest signatures of several Gyr old stars are the red clump (RC) and horizontal branch (HB), both corresponding to core He-burning evolution phases of stars with masses $\lesssim 2$~M$_{\odot}$. In particular, despite the uncertainties on which other parameters actually affect the color extension of the HB (e.g., alpha-element content, internal dynamics, etc.), the potential detection of a blue HB (i.e., stars with $M \lesssim 1$~M$_{\odot}$ at $M_I \simeq 0$ mag) would unambiguously indicate the presence of a population that is both old ($\gtrsim 10$ Gyr) and metal-poor ([Fe/H] $\lesssim -1.5$ dex). Old metal-poor HB stars can cross the instability strip and produce RR Lyrae (RRLs) type pulsating variables, so the detection of such stars allows to unambiguously trace the signature of a $\sim 10$ Gyr old population throughout a galaxy (see, e.g., \citealt{Clementini2003}) even when the magnitude level of the HB is close to the detection limit.

Here we present a detailed analysis of UGC~4483, the closest example of metal-poor BCD galaxy, located between the bright spirals M81 and NGC~2403 at a distance of $D = 3.4 \pm 0.2$~Mpc, corresponding to a distance modulus of DM $= 27.63 \pm 0.12$ \citep{Izotov2002}, and with an oxygen abundance of $12 + \log(O/H) = 7.56 \pm 0.03$ \citep{vanZee2006}, corresponding to $Z \simeq 1/15\ Z_{\odot}$ (using 8.73 for the solar oxygen abundance, \citealt{Caffau2015}). Its cometary shape (see Figure \ref{fig:image}) resembles that of SBS 1415+437 \citep{Aloisi2005}, which also has a very similar metal content. From H\textsc{i} 21 cm observations, \citet{Thuan1979} derived a gas mass of $M_{\textrm{H\textsc{i}}} = 4.1 \times 10^7$ M$_{\odot}$, and a gas fraction which corresponds to 39\% of all visible mass. The H\textsc{i} mass from \citet{Lelli2012b} is $M_{\textrm{H\textsc{i}}} = 2.5 \times 10^7$ M$_{\odot}$, and they also find a steeply-rising rotation curve that flattens in the outer parts, making UGC~4483 the lowest-mass galaxy with a differentially rotating H\textsc{i} disk. This steep rise of the rotation curve indicates a strong central concentration of mass, a property which seems to be typical of BCDs.

\begin{figure*}
\centering
\includegraphics[width=0.8\linewidth]{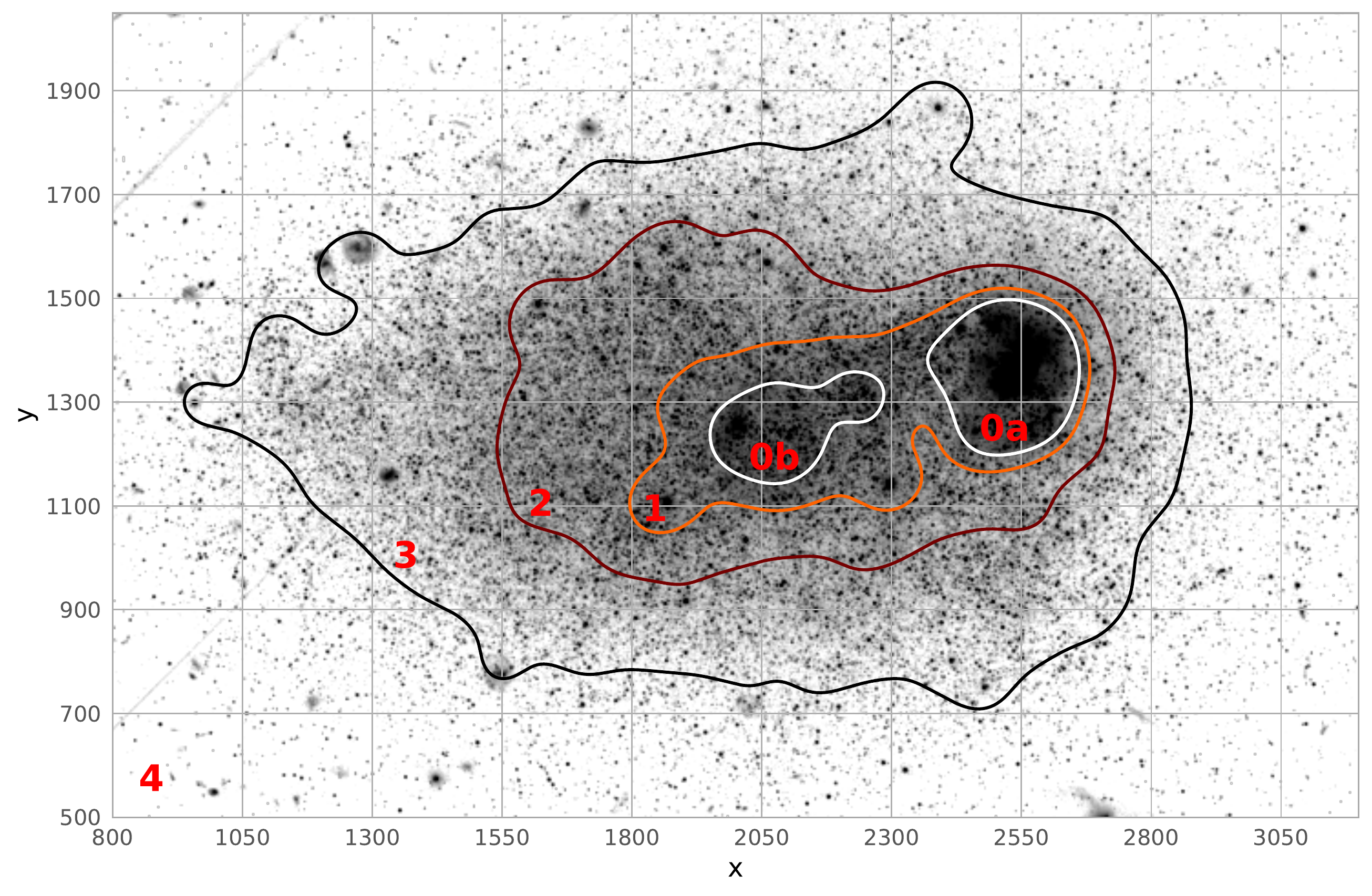}
\caption{F606W image with over-plotted the isophotal contours used to divide the galaxy into different regions (0a, 0b, 1, 2, 3, and 4) and to study the radial behaviour of the stellar populations and SFH. Notice that the image is rotated by 90 degrees to the right with respect to the ones in Figure \ref{fig:image}.}
\label{fig:regions}
\end{figure*}

UGC~4483 has already been resolved into stars with the \textit{HST}/WFPC2 \citep{Dolphin2001,Izotov2002,Odekon2006,McQuinn2010}. The $I$ vs. $V-I$ CMDs reveal a young stellar population of blue MS stars as well as blue and red supergiants associated with the bright H\textsc{ii} region at the northern tip of the galaxy (see Figure \ref{fig:image} and Region 0a of Figure \ref{fig:regions}). An older evolved population was found throughout the whole low surface brightness body of the galaxy, as indicated by very bright AGB stars and the tip of a very blue RGB. However, those data were not deep enough to discriminate between a relatively metal-poor population with an age of $\sim 10$~Gyr for the RGB/AGB stars, or a somewhat higher metallicity and an age of $\sim 2$~Gyr.

We present here new WFC3/UVIS observations reaching more than $4$ mag deeper than the TRGB, to detect and characterize the RC and/or HB populations of the galaxy. 
These new data strongly constrain both the age and metallicity properties of the SFH of UGC~4483 back to many, possibly 10, Gyr ago. The RC absolute magnitude (near $M_I \simeq -0.5$ mag) depends sensitively on the age of the stellar population, and it is $\sim 1$ mag brighter for a 1 Gyr old population than for a 10 Gyr old population (see, e.g., figure 23 of \citealt{Rejkuba2005}). The dependence of the RC magnitude on metallicity is not strong, and either way, is different than the dependence of the RGB color on age and metallicity. Therefore, a joint determination of the RGB color and RC magnitude constrains the age and metallicity of a stellar population independently (see, e.g., NGC 1569 at a similar distance; \citealt{Grocholski2012}). 

The quality of our new data  allows us to finally analyze a very deep CMD of UGC~4483, and to reach lower-mass stars, thus, older stellar features (i.e., the RC and HB) beyond the edge of the Local Group.

\begin{figure}
\centering
\includegraphics[width=\linewidth]{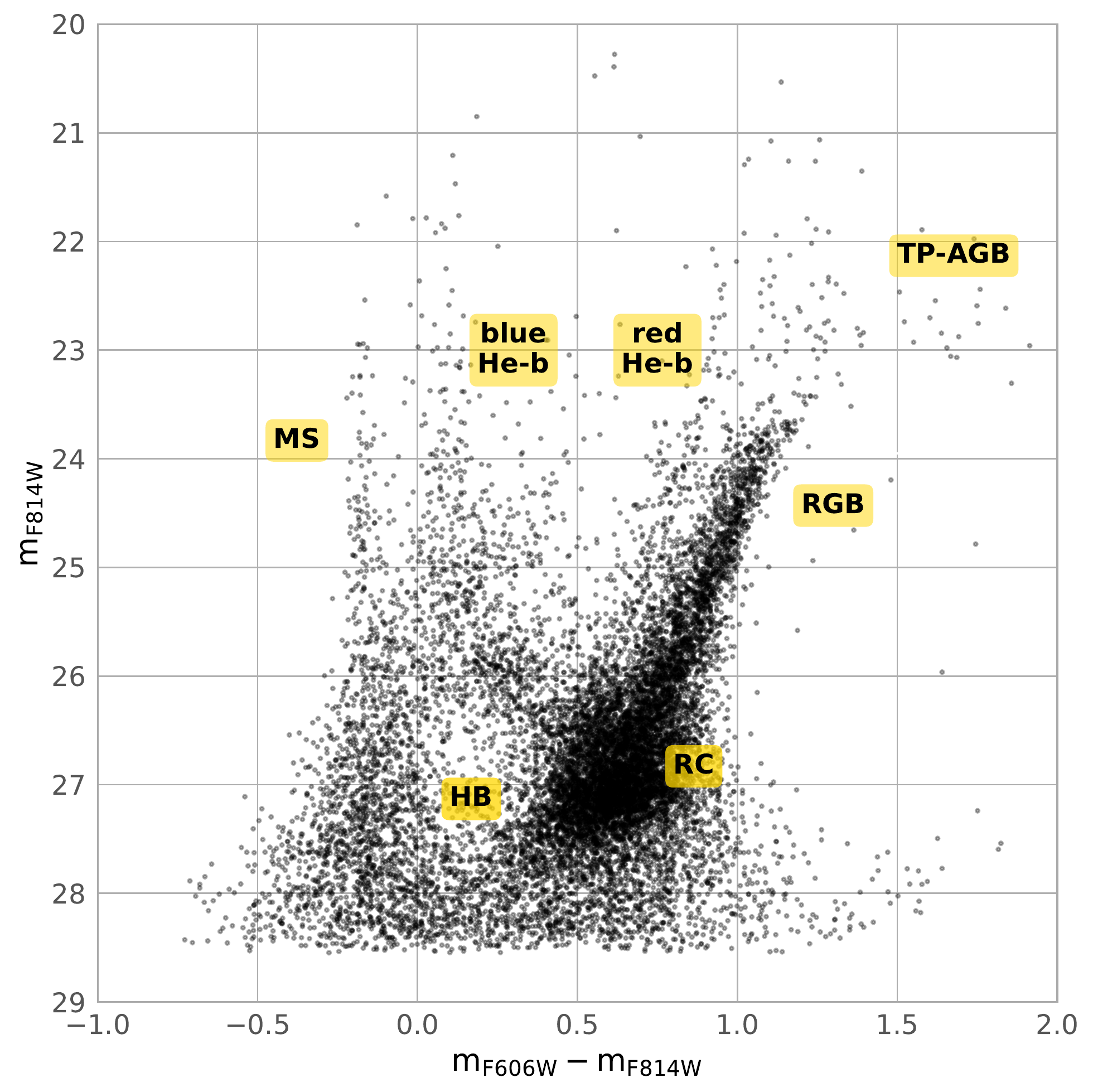}
\caption{F814W versus F606W$-$F814W color-magnitude diagram of UGC~4483 corresponding to the field of view covered by our UVIS imaging (after the quality cuts, see Section \ref{sec:obs}). The main stellar evolutionary phases are indicated (see Section \ref{sec:stellar_pops}).}
\label{fig:cmd}
\end{figure}

\begin{figure*}
\centering
\includegraphics[width=\linewidth]{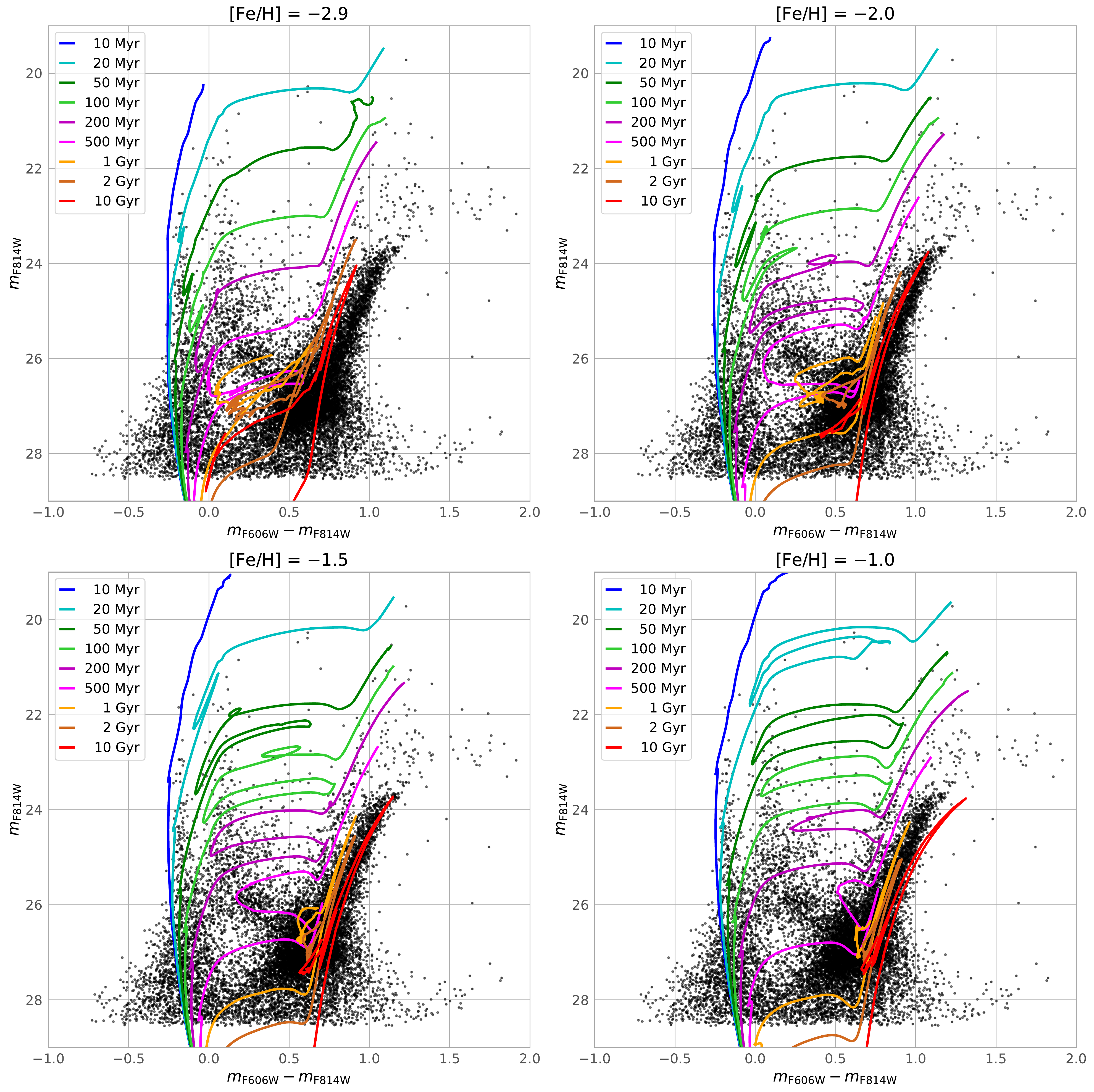}
\caption{Same CMD of Figure \ref{fig:cmd} with MIST isochrones \citep{Choi2016,Dotter2016} of 4 different metallicities over-plotted: [Fe/H] = $-2.9$ (top left), [Fe/H] = $-2.0$ (top right), [Fe/H] = $-1.5$ (bottom left), and [Fe/H] = $-1.0$ (bottom right), shifted to match the distance and foreground extinction of the galaxy. We can recognize the HB feature in both the [Fe/H] = $-2.9$ and the [Fe/H] = $-2.0$ 10 Gyr old isochrones.}
\label{fig:cmd_iso}
\end{figure*}

\begin{figure*}
\centering
\includegraphics[width=\linewidth]{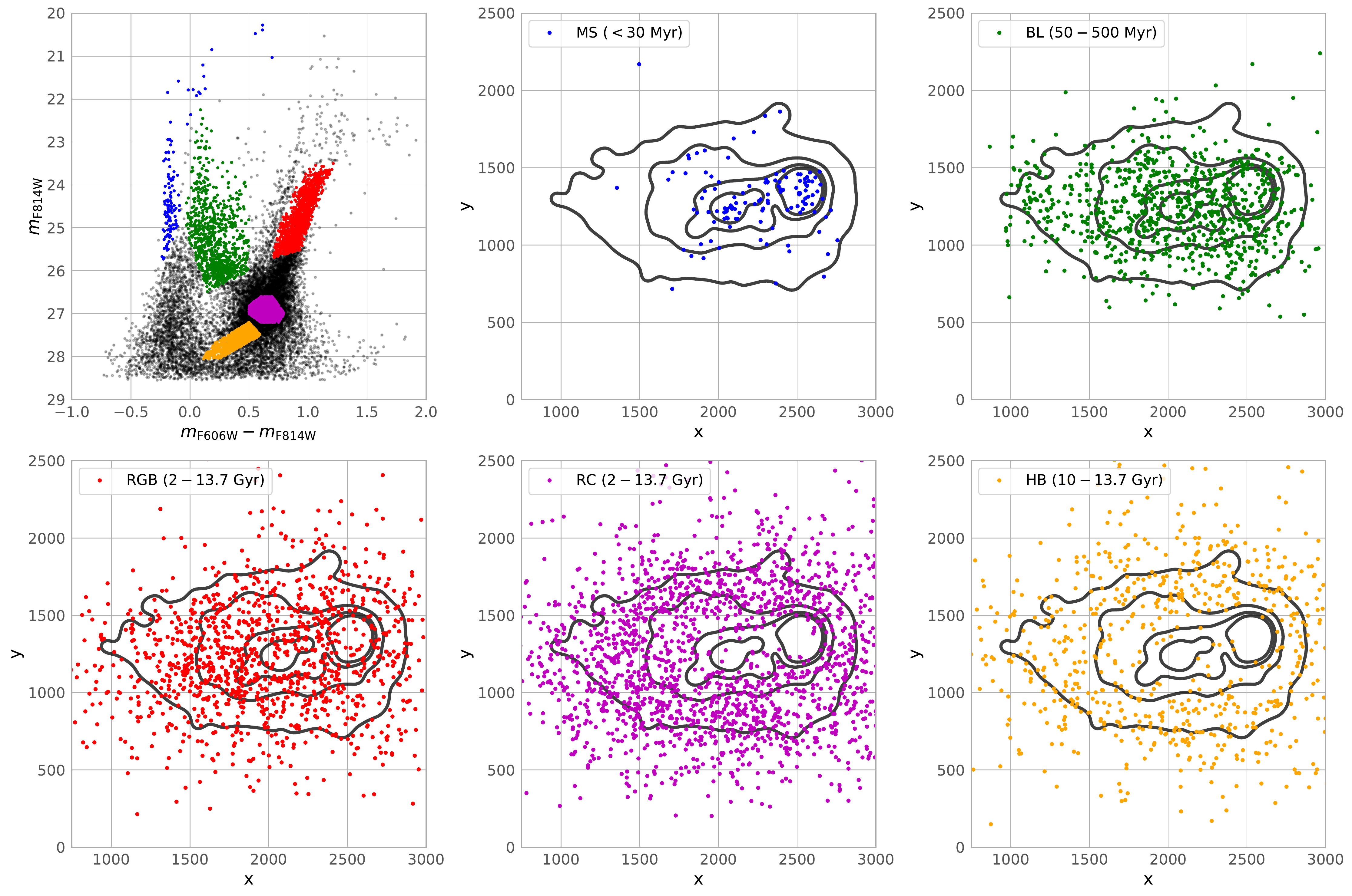}
\caption{\textit{Top left panel.} Selection of different stellar populations in the CMD: in blue, upper MS stars with ages $\lesssim 30$ Myr; in green, BL stars with intermediate ages between $\sim 50$ and  $\sim 500$ Myr; in red, RGB stars older than $\sim 2$ Gyr; in magenta, RC stars older than $\sim 2$ Gyr; in orange, HB stars older than $\sim 10$ Gyr. The other panels show the corresponding spatial distribution of the age-selected stars, as labeled, with contours showing the different regions of the galaxy (see Figure \ref{fig:regions}).}
\label{fig:pops}
\end{figure*}

\section{Observations and Data Reduction} \label{sec:obs}

Observations of UGC~4483 were performed on January 2019 (Visits 1 and 2) and February 2019 (Visit 3) using the UVIS channel of the WFC3 as part of the {\it HST} program GO-15194 (PI: A. Aloisi). Despite the smaller field of view, we preferred WFC3/UVIS over ACS/WFC because of the slightly higher resolution that provides a better photometry in the most crowded central regions of the galaxy. The target was centered on one of the two CCD chips of the WFC3/UVIS camera, in order to minimize the impact of chip-dependent zeropoints and to avoid the loss of the galaxy central region due to the CCD chip gap. The observations were obtained in the two broad-band filters F606W and F814W. We also required narrow-band imaging in F656N to study the distribution of the ionized gas. We selected the broader F606W instead of F555W as the blue filter, because it provides the best compromise between a reasonable exposure time (a factor of $\sim 2$ shorter for F606W than for F555W) and the achievement of our science goals. During Visits 1 and 2, 14 long exposures (2580 sec each) were taken in the F606W filter (for a total exposure time of 36120 sec), and 8 in the F814W filter (for a total exposure time of 20640 sec). Twelve additional exposures with the same integration time (2580 sec each, for a total exposure time of 30960 sec) were collected during Visit 3 in the F814W filter. Those exposure times were chosen in order to achieve a signal-to-noise ratio of $\sim 10$, at the HB magnitude level, which corresponds to $I \sim 27.7$ mag at the distance of UGC~4483. The exposures were executed with a spatial offset, using a dithering pattern of an integer+fraction of pixels, in order to move across the gap between the two chips, to simplify the identification and removal of bad/hot pixels and cosmic rays, and to improve the point spread function (PSF) sampling.

It is worth mentioning that the observations were performed in the Continuous Viewing Zone (CVZ) of {\it HST}, which allowed us to observe for the entire 96 minute orbit. This resulted in a very deep photometry even in just 18 orbits.

To reduce the images, we followed the same procedure outlined in \citet{Annibali2019}. First, the calibrated \texttt{.flc} science images were retrieved from the {\it HST} archive. The \texttt{.flc} images are the products of the \texttt{calwf3} data reduction pipeline and constitute the bias-corrected, dark-subtracted, flat-fielded, and charge transfer inefficiency corrected images. Then, we combined together the individual \texttt{.flc} images into a single drizzled, stacked, and distortion-corrected image (\texttt{.drc} image) using the \texttt{Drizzlepac} software \citep{Gonzaga2012}. To do so, all the images in the same filter were first aligned using the software \texttt{TweakReg} and then, using the software \texttt{AstroDrizzle}, bad pixels and cosmic rays were flagged and rejected from the input images. Finally, those input undistorted and aligned images were combined together into a final stacked image. 

In the left panel of Figure \ref{fig:image} we show a 3-color composite image of UGC~4483 from our \textit{HST}/WFC3 observations: blue corresponds to F606W (broad $V$), red to F814W ($I$), while the green channel was obtained using the mean of the two. The right panel shows instead the F656N image of the galaxy (please note that given its very low S/N, the H$\alpha$ image has not been continuum subtracted.) H\textsc{i} contours from the VLA-ANGST survey \citep{Ott2012} corresponding to column densities of $N_H=$ 3.0, 2.5, 2.0, 1.5, 1.0, and $0.5 \times 10^{21}$ cm$^{-2}$ have been superimposed to both images. Nebular gas emission can contaminate broad-band photometry, but as shown by the right panel of Figure \ref{fig:image}, outside of the most active 0a and 0b regions (as defined in Figure \ref{fig:regions}), the contamination is negligible.

PSF stellar photometry was performed using the latest version of Dolphot (\citealt{Dolphin2000}, and numerous subsequent updates). After the usual pre-processing required by the software and performed on each single science image (i.e., creation of bad pixel mask and generation of sky frame), iterative PSF photometry was performed simultaneously on the \texttt{.flc} images using the F606W \texttt{.drc} image as the reference frame for alignment. The different Dolphot parameters that govern alignment and photometry were set as in \citet{Annibali2019}, adopting a hybrid combination between the values recommended by the Dolphot manual and those adopted in \citet{Williams2014}. 

Together with the positions and photometry of the individual stars, the final photometric catalog contains several diagnostic parameters which are useful to exclude remaining artifacts and spurious detections. Hence, we selected from the total catalog all the objects with the Dolphot ``object type'' flag = 1, and then we applied a series of consecutive selection cuts using the remaining diagnostics (i.e., error, sharpness, roundness, and crowding). The final clean catalog contains $\sim 14\,000$ stars, and the corresponding CMD is shown in Figure \ref{fig:cmd}.

We also analyzed the coordinated parallel images obtained with the ACS/WFC, which in principle could give us rich information about the stellar population of the halo of the galaxy. However, the resulting CMD does not contain any sources we could link to UGC~4483, but only background contamination; unfortunately, these fields are probably already too far away from the galaxy to include its halo. This suggests that the halo of UGC~4483 does not reach as far as the ACS field, i.e. is smaller than $\sim 6$~kpc.

To properly analyze the spatial variations of the SF within the galaxy, we divided our final catalog in six sub-regions, as shown in Figure \ref{fig:regions}, following the isophotal contours of the F606W image. The innermost regions, 0a and 0b, correspond to the most active areas of the galaxy, where we see H$\alpha$ emission from the H\textsc{ii} regions.


\section{Stellar populations} \label{sec:stellar_pops}
Figure \ref{fig:image} clearly shows the very young stellar population of this BCD galaxy. The bright H\textsc{ii} region at the northern edge of UGC~4483 is evident in both the composite 3-color image in the left panel and in H$\alpha$ emission in the right panel, which also reveals another active region more to the south. The H\textsc{i} distribution, shown with the contours overlapped on the images, does not exactly follow the H$\alpha$, which can be due to the fact that the recently formed massive stars ionized the gas (thus we see H$\alpha$ emission) leaving less neutral gas in their surroundings (and also external effects like a recent interaction might have influenced the H\textsc{i} gas distribution).

Figure \ref{fig:cmd} shows the F814W versus F606W$-$F814W CMD of UGC 4483. We notice the extremely high quality of these data, with the main stellar evolutionary phases clearly recognizable and well separated in the diagram. The CMD shows a continuity of stellar populations, suggesting a SF that spans from ancient to recent epochs; we can identify two well separated sequences for MS stars at $\mathrm{m_{F606W}-m_{F814W}} \lesssim 0$ and stars at the blue edge of the BLs (core He-burning intermediate- and high-mass stars) at $\mathrm{m_{F606W}-m_{F814W}} \sim 0.2$, the diagonal feature produced by stars at the red edge of the BLs at $\mathrm{m_{F606W}-m_{F814W}} \gtrsim 0.5$ (well separated from the RGB), and a few carbon stars and thermally pulsing asymptotic giant branch (TP-AGB) stars, outlining the horizontal feature around $\mathrm{m_{F606W}-m_{F814W}} \gtrsim 1$ and $\mathrm{m_{F814W}} \lesssim 23$; older stars are enclosed in the RGB with its clearly defined tip at $\mathrm{m_{F814W}} \sim 23.7$, in the RC, and possibly in the HB, our oldest age signature.

To guide the eye and explore the age and metallicity of these populations, we over-plot the MIST isochrones \citep{Choi2016,Dotter2016} for four different metallicities ([Fe/H] = $-2.9$, $-2.0$, $-1.5$, $-1.0$) on the CMD (Figure \ref{fig:cmd_iso}). As a reference, from the oxygen abundance of the H\textsc{ii} regions we obtain a current metallicity [Fe/H] = $-1.2$. The isochrones were shifted according to galaxy's distance modulus of DM = 27.45 (which is what we obtain from out best-fitting technique, see Section \ref{sec:sfh}), and foreground extinction of $\mathrm{E(B-V)}=0.034$ \citep{Schlafly2011}. The different metallicities are labeled at the top of each plot, while ages are labeled in the legend. We can see how the most metal-poor isochrone set is too blue to fit the youngest stellar populations, in particular to reproduce the color of the BLs, which instead starts to be compatible with a metallicity of [Fe/H]~=~$-1.5$. Also RGB stars are significantly redder than the two most metal-poor isochrone sets shown here, even though they touch its blue edge. We cannot instead exclude such low metallicities for older, fainter stars.

To understand how the different stellar populations are distributed within the galaxy, we select age intervals in the CMD using the same MIST isochrones as a guide, and plot the spatial map of the selected stars. The results are shown in Figure \ref{fig:pops}. We choose to select upper MS stars (in blue), with ages $\lesssim 30$ Myr, blue BL stars (in green), with intermediate ages between $\sim 50$ and  $\sim 500$ Myr, RGB stars (in red), older than $\sim 2$ Gyr, RC stars (in magenta), older than $\sim 2$ Gyr, and HB stars (in orange), older than $\sim 10$ Gyr. The top left panel shows these selections in the CMD, while in the other panels we plot the spatial maps of the age-selected stars, as labeled, with contours showing the different regions we selected within the galaxy (see Figure \ref{fig:regions}).

The general trend is that younger stars have more concentrated and clumpy distributions compared to older stars, a behaviour expected as stars move out of their natal structures as they age.
In particular, very young MS stars are mainly found in the two inner regions (0a and 0b) which host the very bright SF regions visible in Figures \ref{fig:image} and \ref{fig:regions}, where the H\textsc{ii} regions are located. We also notice the central holes in these same regions in the RGB, RC, and HB maps, a sign of the strong incompleteness of these fainter stellar populations there. In reality, these faint old stars are likely uniformly distributed over the galaxy.



\section{Artificial Star Tests} \label{sec:ast}

To properly characterize the photometric errors and incompleteness of the data, we perform artificial star tests (ASTs) on our images using the dedicated DOLPHOT routine.

We follow a general standard procedure where we add a fake star (for which we know exactly the input position and magnitudes) to the real images, re-run the photometry, and check whether the source is detected and its output magnitudes. We then repeat the process many times (2 millions in this case) varying the position and magnitudes of the input fake stars, to fully map the whole image and explore the whole range of magnitudes and colors covering the observed CMD. We inject each individual star simultaneously in both F606W and F814W images, in order to reproduce a realistic situation and to account for the error and completeness correlation in the two bands. We consider a star ``recovered'' in the output catalog if the measured magnitude is within 0.75 mag from its input value in both filters, and satisfy all the selection cuts applied to the real photometry. Adding one fake star at a time guarantees not to artificially alter the crowding of the images.

Given the very different crowding conditions within the galaxy, we build the input distribution of stars following the surface brightness of the F606W image, to obtain a more accurate estimate of the incompleteness in the most crowded regions that would be under-sampled with a uniform input distribution.

\begin{figure}
\centering
\includegraphics[width=\linewidth]{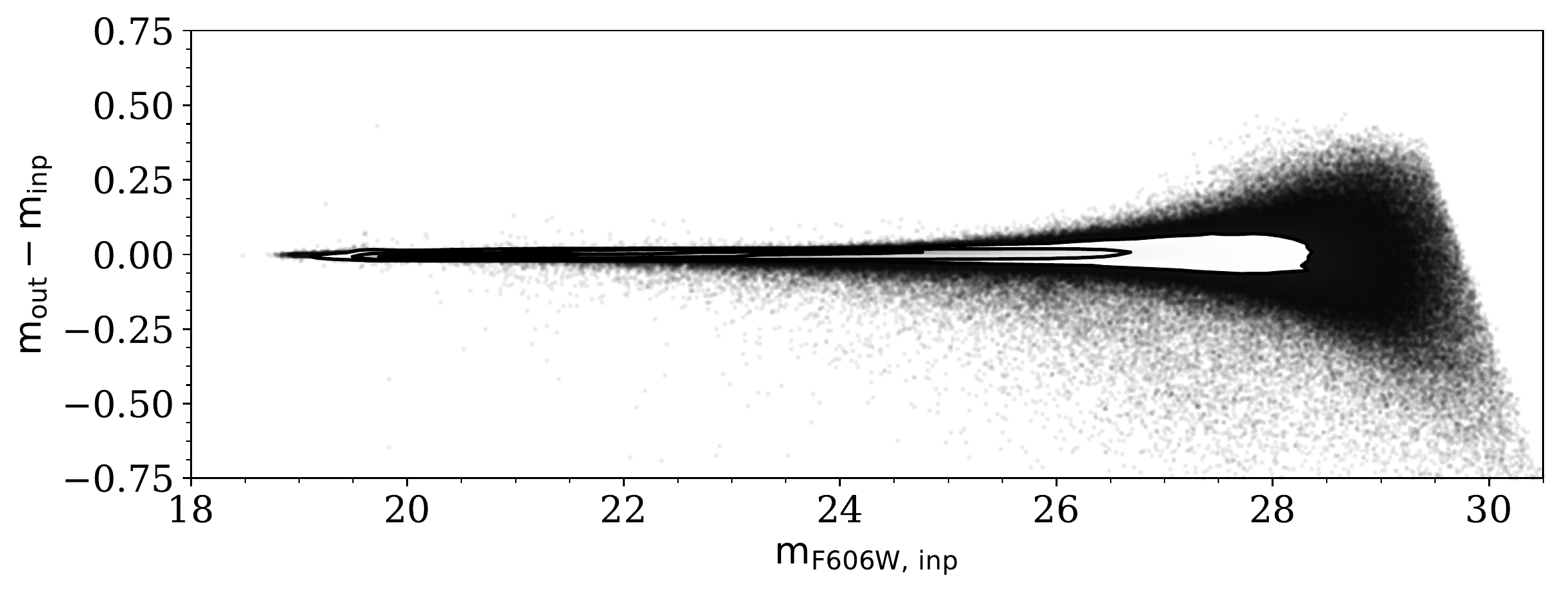}
\centering
\includegraphics[width=\linewidth]{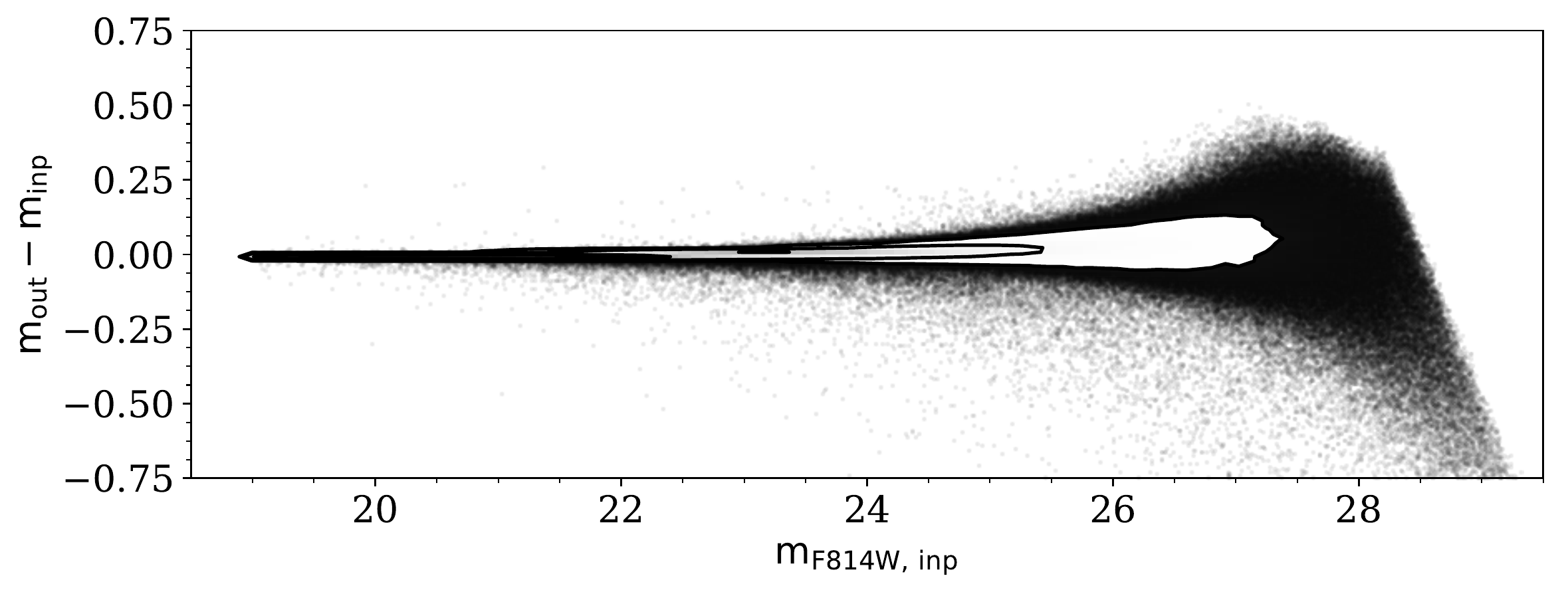}
\caption{Photometric errors in F606W (top panel) and F814W (bottom panel) from our artificial star tests; the contours indicate the 1$\sigma$, 2$\sigma$ and 3$\sigma$ levels of the distributions.}
\label{fig:ast}
\end{figure}

\begin{figure}
\centering
\includegraphics[width=\linewidth]{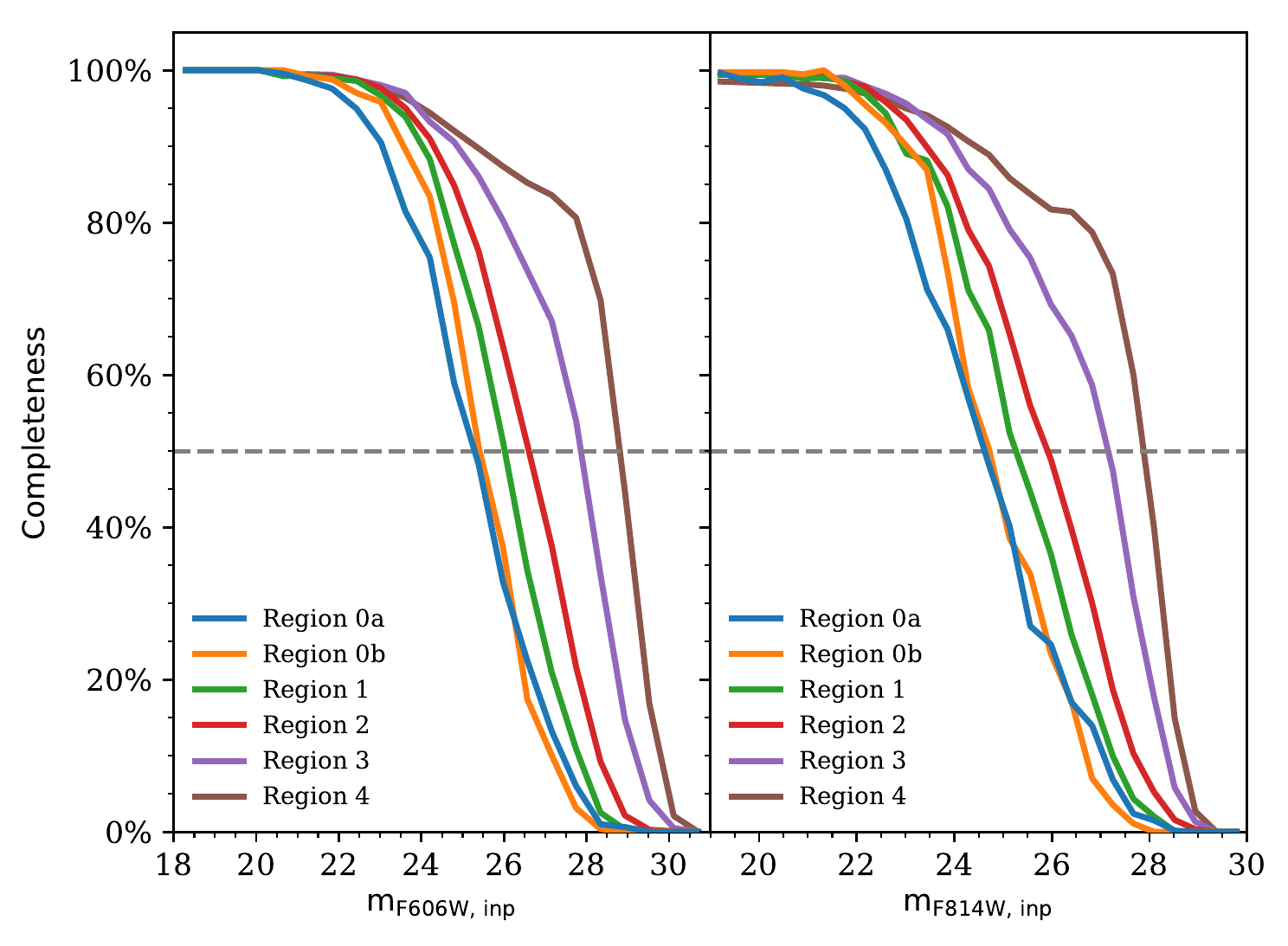}
\caption{Completeness in F606W (left panel) and F814W (right panel) from our artificial star tests in the various regions of the galaxy, highlighting the very different crowding conditions from inside out. The dashed horizontal line marks the 50\% completeness level.}
\label{fig:completeness}
\end{figure}

From the output distribution of the recovered stars we derive an estimate of the photometric error (from the $m_{output}-m_{input}$ versus $m_{input}$ distribution, Figure \ref{fig:ast}) and completeness (from the ratio between the number of output and input stars, Figure \ref{fig:completeness}) as a function of both space and magnitude. 
Generally speaking, a star is considered recovered if its output flux agrees within 0.75 mag with its input value. We also consider ``lost'' the stars that do not pass the same quality tests as the real data. The distribution of the photometric error displayed in Figure \ref{fig:ast} also allows to take into account the effect of blending of multiple sources on the photometry; in fact, a systematic negative skewness in the output$-$input flux (stars found brighter than their input) is a signature of overlapping of artificial stars with other ones. We take this skewness into account to consider the blending effect when we create the synthetic CMDs.

\begin{figure*}
\centering
\includegraphics[width=\linewidth]{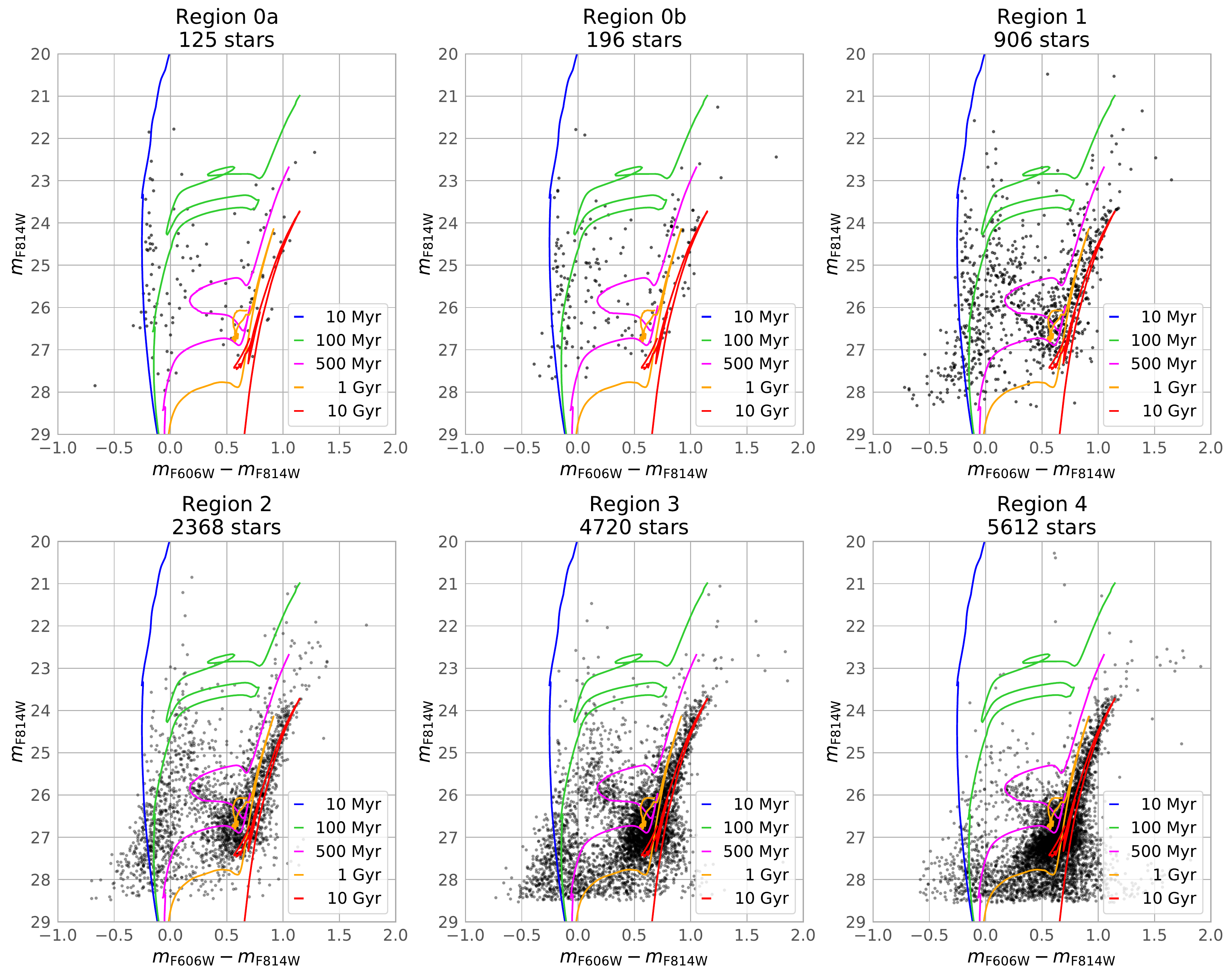}
\caption{CMDs of the regions we identified in the galaxy (see Figure \ref{fig:regions}) and used to recover the SFH (0a and 0b being the innermost ones, 4 the outermost one, see Figure \ref{fig:regions}). Over-plotted in colors are the MIST isochrones \citep{Choi2016,Dotter2016} with [Fe/H] = $-1.5$ and ages as labeled, shifted to match the distance and foreground extinction of the galaxy.}
\label{fig:cmd_regions}
\end{figure*}

In Figure \ref{fig:completeness} we plot the completeness as a function of magnitude for the different regions in which we divided the galaxy (see Figure \ref{fig:regions}). The very different crowding conditions from inside out are reflected in the different completeness behaviours: indeed, the most internal regions are the most crowded, thus incomplete ones, while the completeness increases as we move outwards. The completeness is 50\% at $\mathrm{m_{F606W}} \simeq 25.3$ and $\mathrm{m_{F814W}} \simeq 24.75$ in the central regions of the galaxy (0a and 0b), and at $\mathrm{m_{F606W}} \simeq 28.7$ and $\mathrm{m_{F814W}} \simeq 28$ in the outer field (Region 4). 


\section{Star Formation History} \label{sec:sfh}
With the photometric catalog and the artificial star catalog in our hands, we can perform detailed studies of the stellar populations and recover the SFH of UGC~4483 using the synthetic CMD method \citep[see, e.g.,][]{Tosi1991,Gallart2005,Tolstoy2009,McQuinn2010,Weisz2011,Cignoni2015,Sacchi2018}: the observed CMD is compared to synthetic ones built from a set of stellar evolution models (evolutionary tracks or isochrones) adequately treated to match the distance, extinction, and photometric properties of the galaxy. Synthetic CMDs, each representing a simple stellar population of fixed age and metallicity, are created and used as ``basis'' functions (BFs), and a linear combination of these BFs creates a composite population which can represent, with the appropriate weights, any SFH. The weight associated with each BF is proportional to the number of stars formed at that age and metallicity, and the best-fit SFH is described by the set of weights producing a composite model CMD most similar to the observed one. The best-fitting weights are determined by using a minimization algorithm to compare data and models.

We build our models from both the PARSEC-COLIBRI \citep{Bressan2012,Marigo2017} and MIST \citep{Choi2016,Dotter2016} isochrone libraries\footnote{Notice that the PARSEC-COLIBRI models adopt $Z_{\odot} = 0.0152$, while the MIST models adopt $Z_{\odot} = 0.0142$.} using the following parameters: Kroupa initial mass function \citep{Kroupa2001} from 0.1 to 350 M$_{\odot}$; 30\% binary fraction; [Fe/H] from $-2.9$ to $1.0$ in steps of $0.1$; DM = 27.60 \citep{Izotov2002}; foreground extinction $\mathrm{E(B-V)}=0.034$ \citep{Schlafly2011}; photometric errors and incompleteness from our ASTs (see Section \ref{sec:ast}, and Figures \ref{fig:ast} and \ref{fig:completeness}). We also explore different distance/reddening combinations to obtain the best match with the data.

We do not impose a metallicity distribution function, but we require that the metallicity at each time bin cannot be more than 25\% lower than the metallicity of the adjacent older bin. This is to include a soft constraint without forcing an unknown distribution \textit{a priori}. This is a looser boundary condition with respect to a metallicity monotonically increasing with time, often imposed in this kind of studies; however, in a galaxy as metal-poor as UGC~4483, nobody knows exactly how the metallicity varies with time, for instance because of the accretion of large amounts of metal poor gas, or because of significant losses of heavy elements, through galactic winds triggered by powerful supernova explosions.

We use the hybrid genetic code SFERA \citep{Cignoni2015} to build the synthetic CMDs and perform the comparison between models and data. SFERA is built to performs a complete implementation of the synthetic CMD method, from the MonteCarlo extraction of synthetic stars from the adopted stellar evolution models, to the inclusion of  observational effects in the models through ASTs, and finally, performing the minimization of the residuals between observed and synthetic CMDs, with an accurate estimate of the uncertainties. For the minimization, we bin both the synthetic and observed CMDs to compare star counts in each grid cell. Given that we need to take into account the possible low number counts in some CMD cells, we choose to follow a Poissonian statistics, looking for the combination of synthetic CMDs that minimizes a likelihood distance between model and data:
\begin{eqnarray}  
\chi_P= \sum_{i=1}^{N
    bin} obs_{i}\ln\frac{obs_{i}}{mod_{i}}-obs_{i}+mod_{i}
\label{pois1} 
\end{eqnarray}
where $mod_{i}$ and $obs_{i}$ are the model and the data histograms in the $i-$bin. This likelihood is minimized with the hybrid-genetic algorithm, i.e. a combination of a classical genetic algorithm (Pikaia\footnote{Routine developed at the High Altitude Observatory and publicly available: \url{http://www.hao.ucar.edu/modeling/pikaia/pikaia.php}}) and a local search (Simulated Annealing).

Statistical uncertainties are computed with a bootstrap technique on the data, i.e. applying small shifts in color and magnitude to the observed CMD and re-deriving the SFH for each new version. We average the different solutions and take the rms deviation as statistical error on the mean value. Systematic uncertainties are accounted for by re-deriving the SFH with different grids to bin age and CMDs, and using different sets of isochrones. These are added in quadrature to the statistical error.

Before discussing the details of the SFH in the various regions, we can have a look at their CMDs which already reveal profound differences within the body of the galaxy. Figure \ref{fig:cmd_regions} shows the 6 CMDs of these regions (0a and 0b being the innermost ones, 4 the outermost one, see Figure \ref{fig:regions}), with over-plotted in colors the MIST isochrones with [Fe/H] = $-1.5$ and ages in the range 10 Myr $-$ 10 Gyr as labeled, shifted to match the distance and foreground extinction of the galaxy.

Regions 0a and 0b include the most actively star forming areas of the galaxy, where we see H$\alpha$ emission from the H\textsc{ii} regions. Their CMDs show very young ($\lesssim 10$~Myr) MS stars, with some BL stars, and very few RGB stars. The observed MS in these two regions is a bit redder with respect to the plotted isochrones, and presumably needs a higher metallicity and/or larger reddening, which is most likely in these highly star-forming regions. As already pointed out in Section \ref{sec:stellar_pops}, the presence of very bright young stars makes the incompleteness severe here, hindering a proper characterization of all the stellar populations present in these regions. This, together with the low number of stars observed, prevents us from running our SFH procedure in a statistically significant way.

\begin{figure*}
\centering
\includegraphics[width=0.55\linewidth]{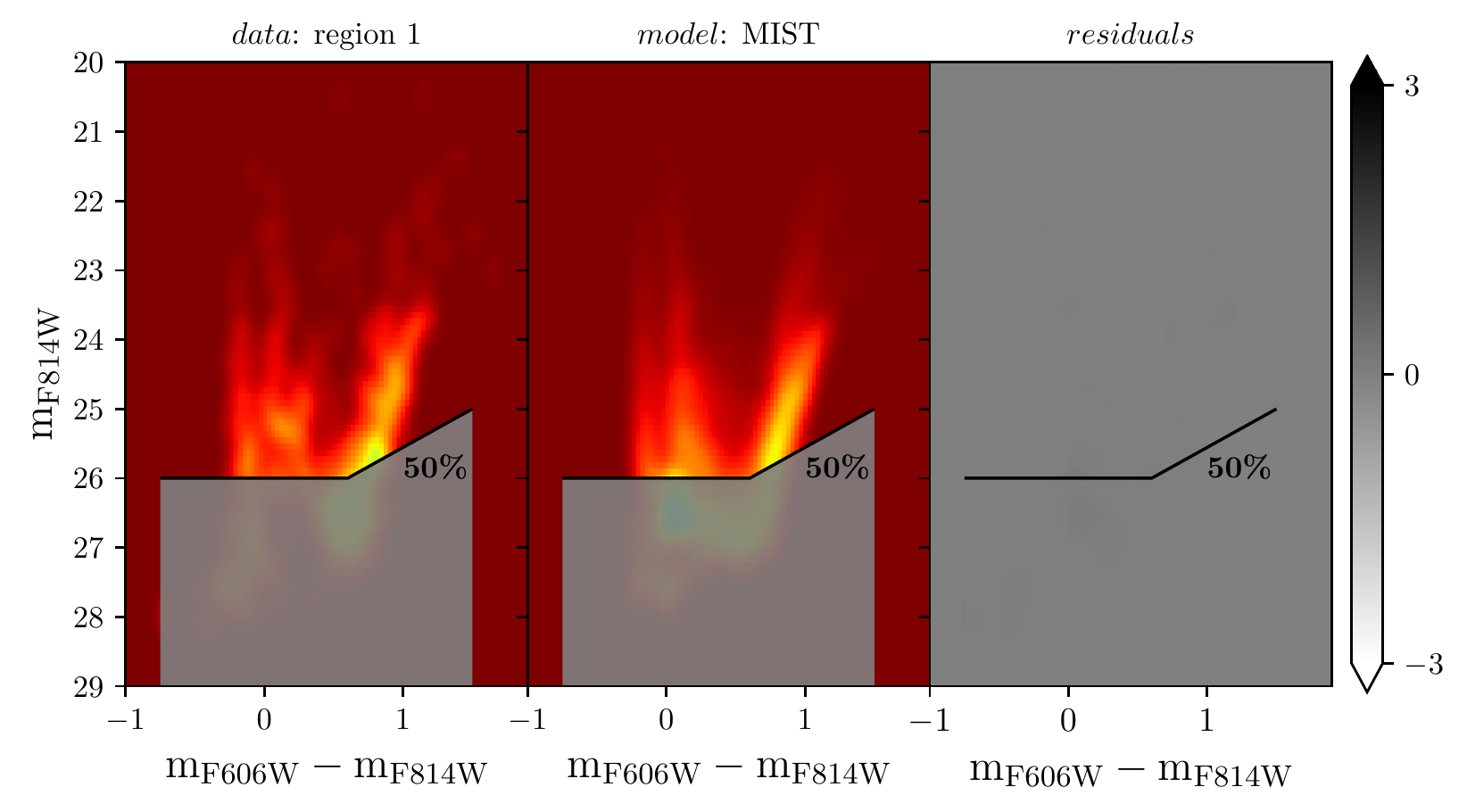}
\hspace{0.5cm}
\includegraphics[width=0.36\linewidth]{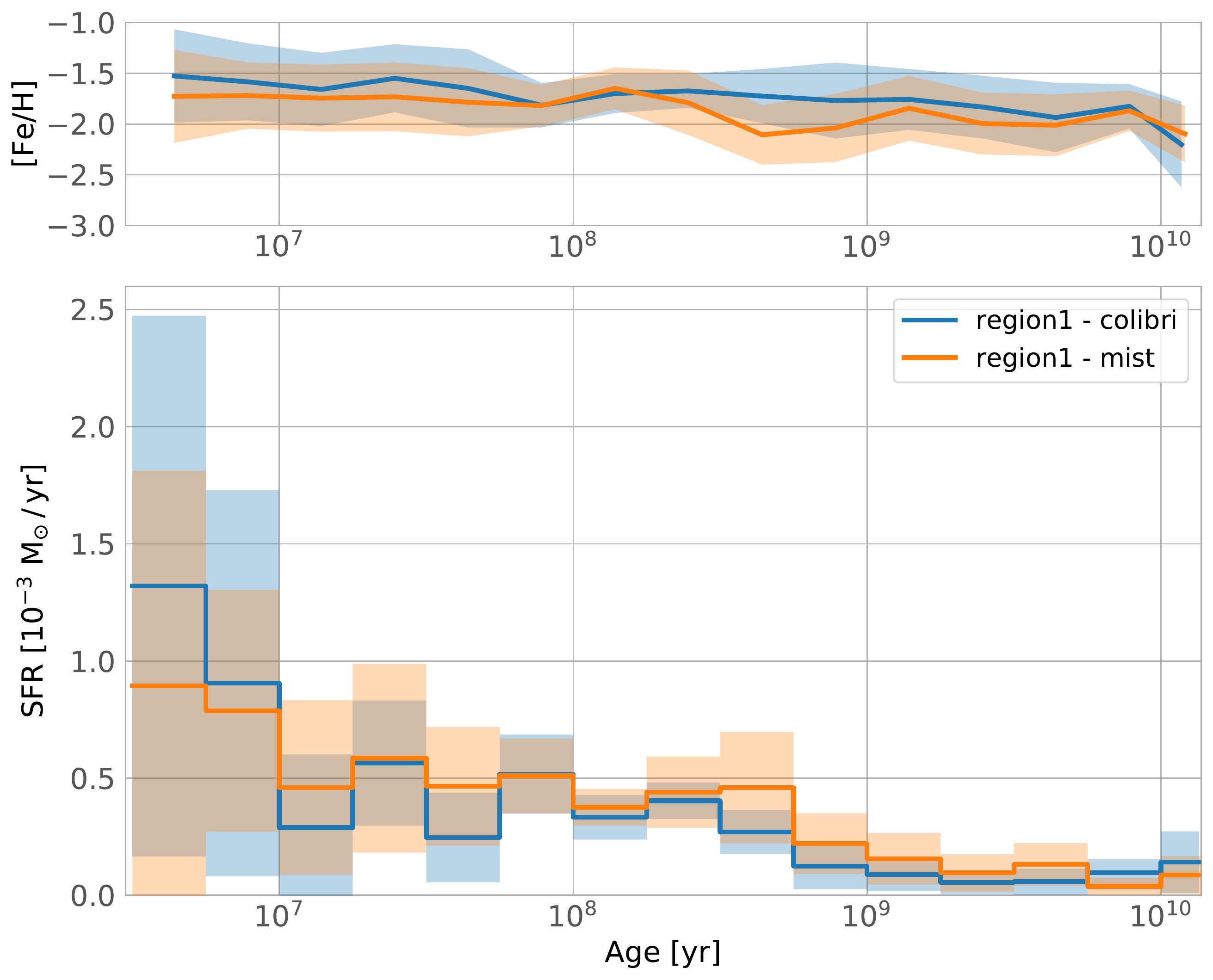}

\includegraphics[width=0.55\linewidth]{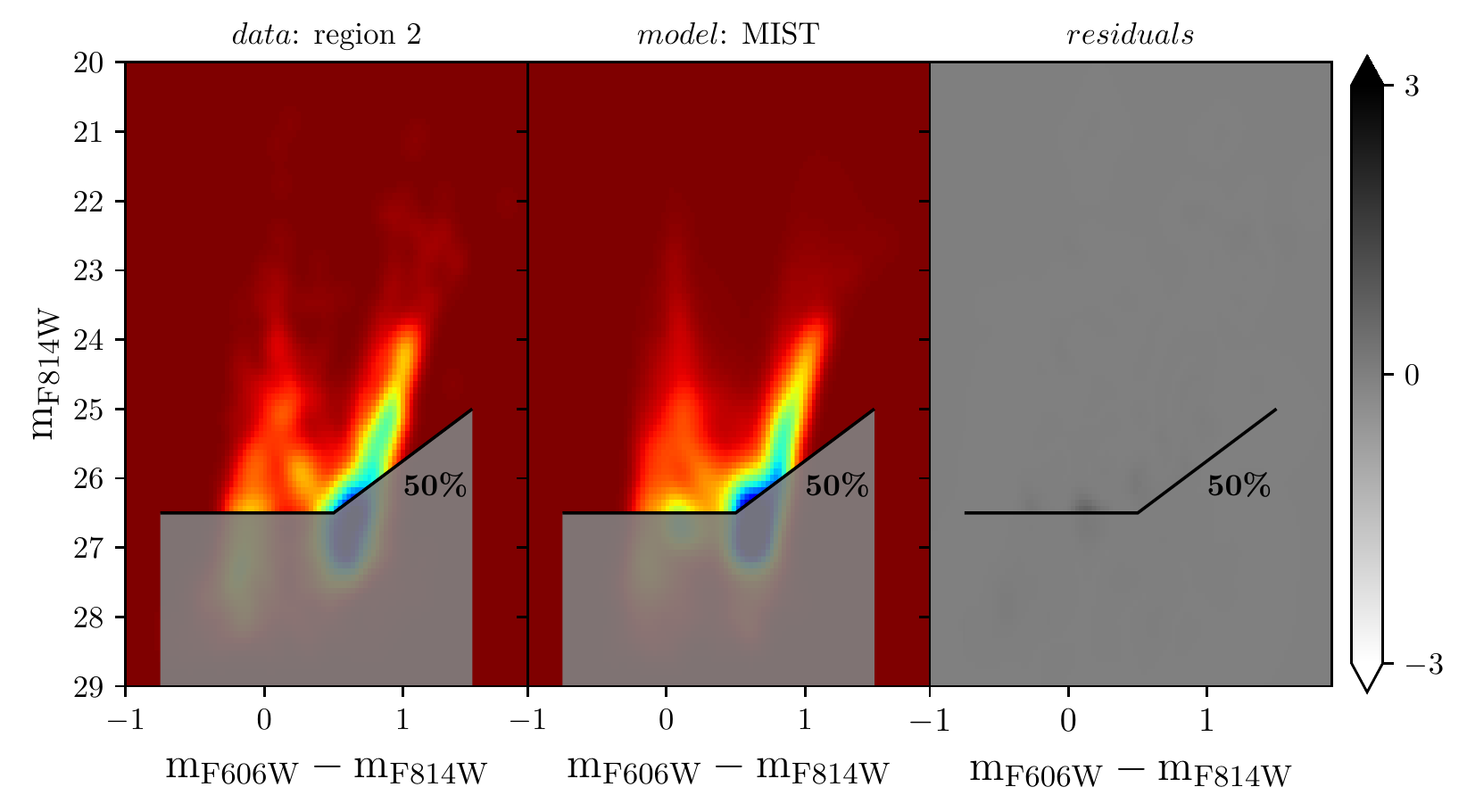}
\hspace{0.5cm}
\includegraphics[width=0.36\linewidth]{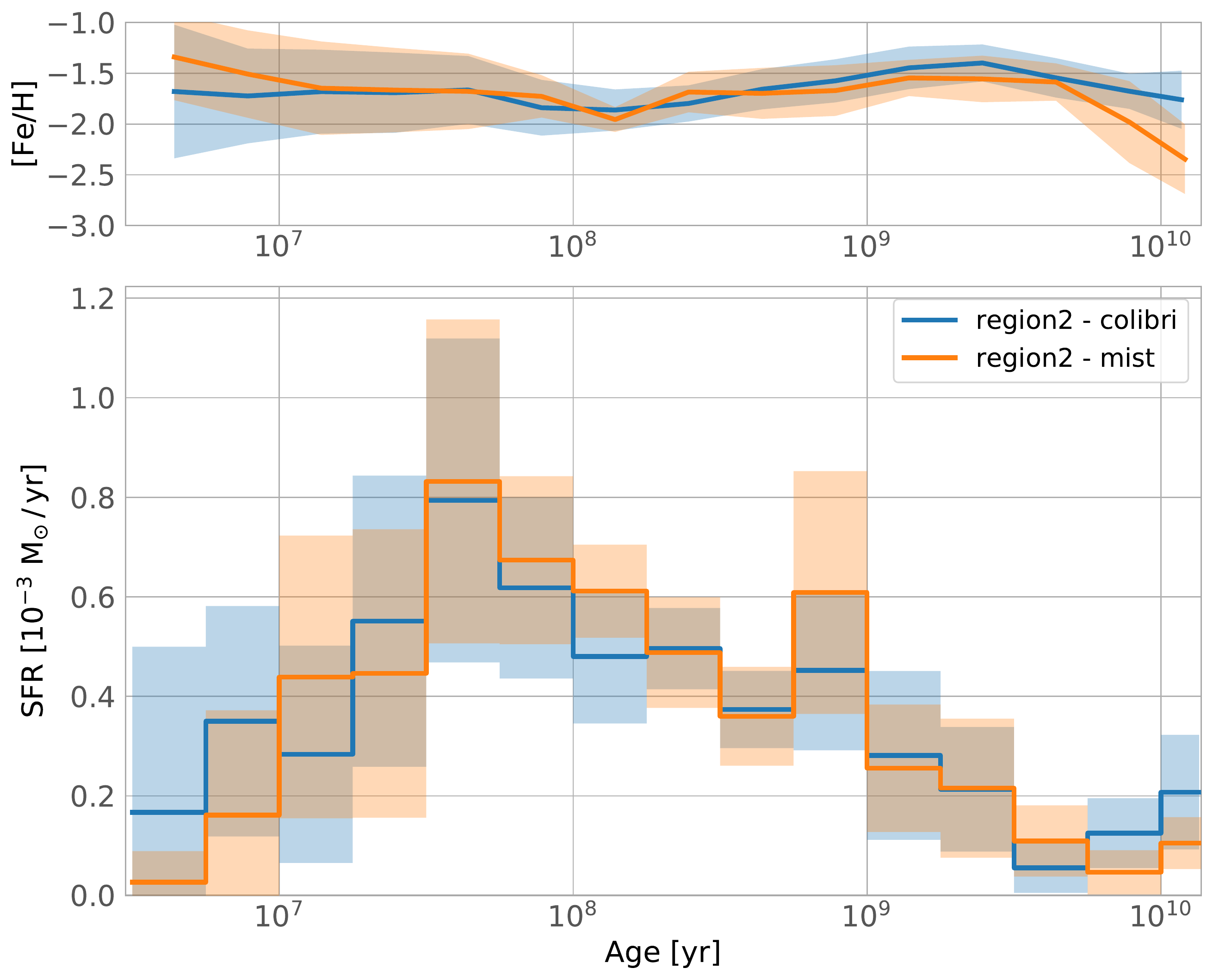}

\includegraphics[width=0.55\linewidth]{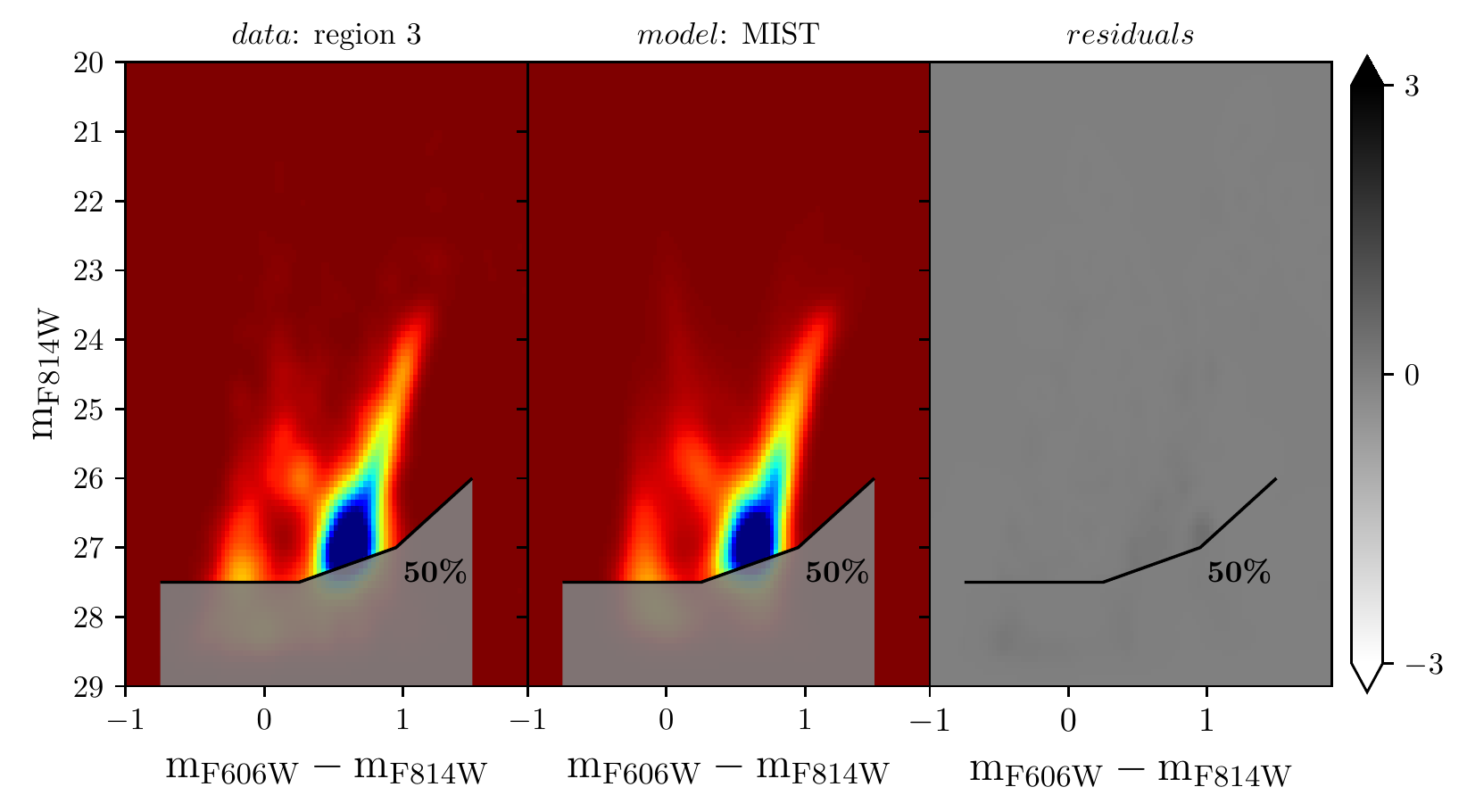}
\hspace{0.5cm}
\includegraphics[width=0.36\linewidth]{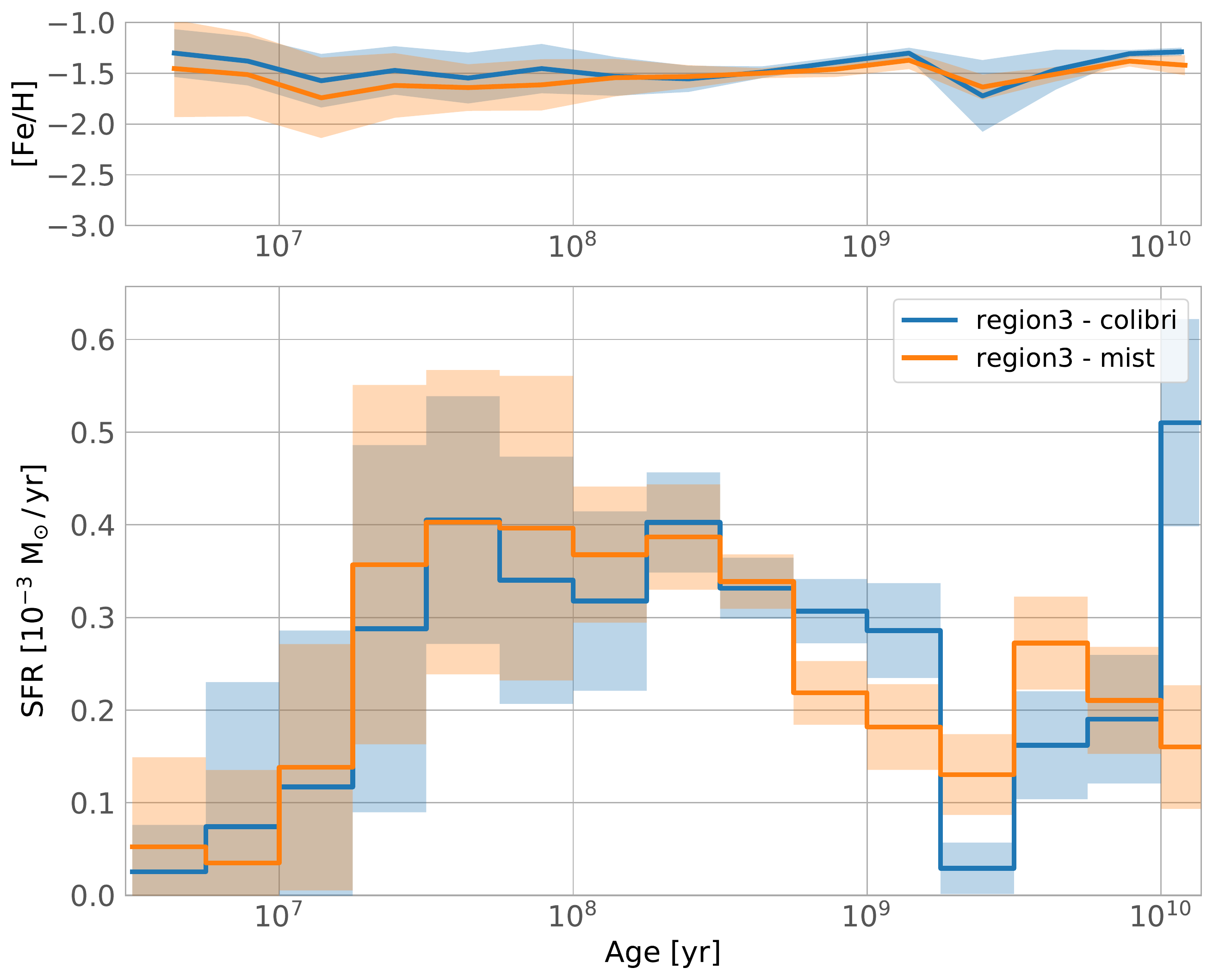}

\includegraphics[width=0.55\linewidth]{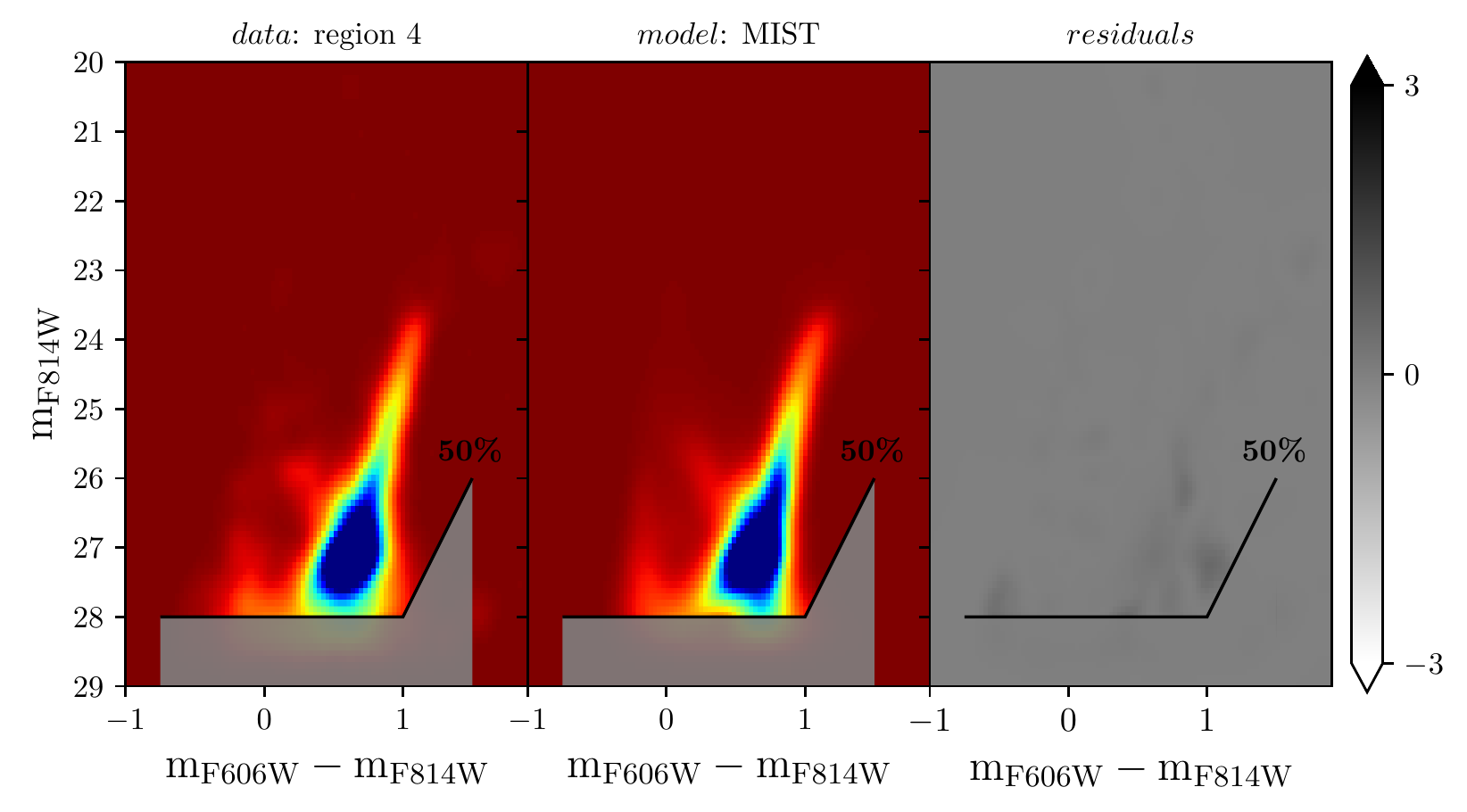}
\hspace{0.5cm}
\includegraphics[width=0.36\linewidth]{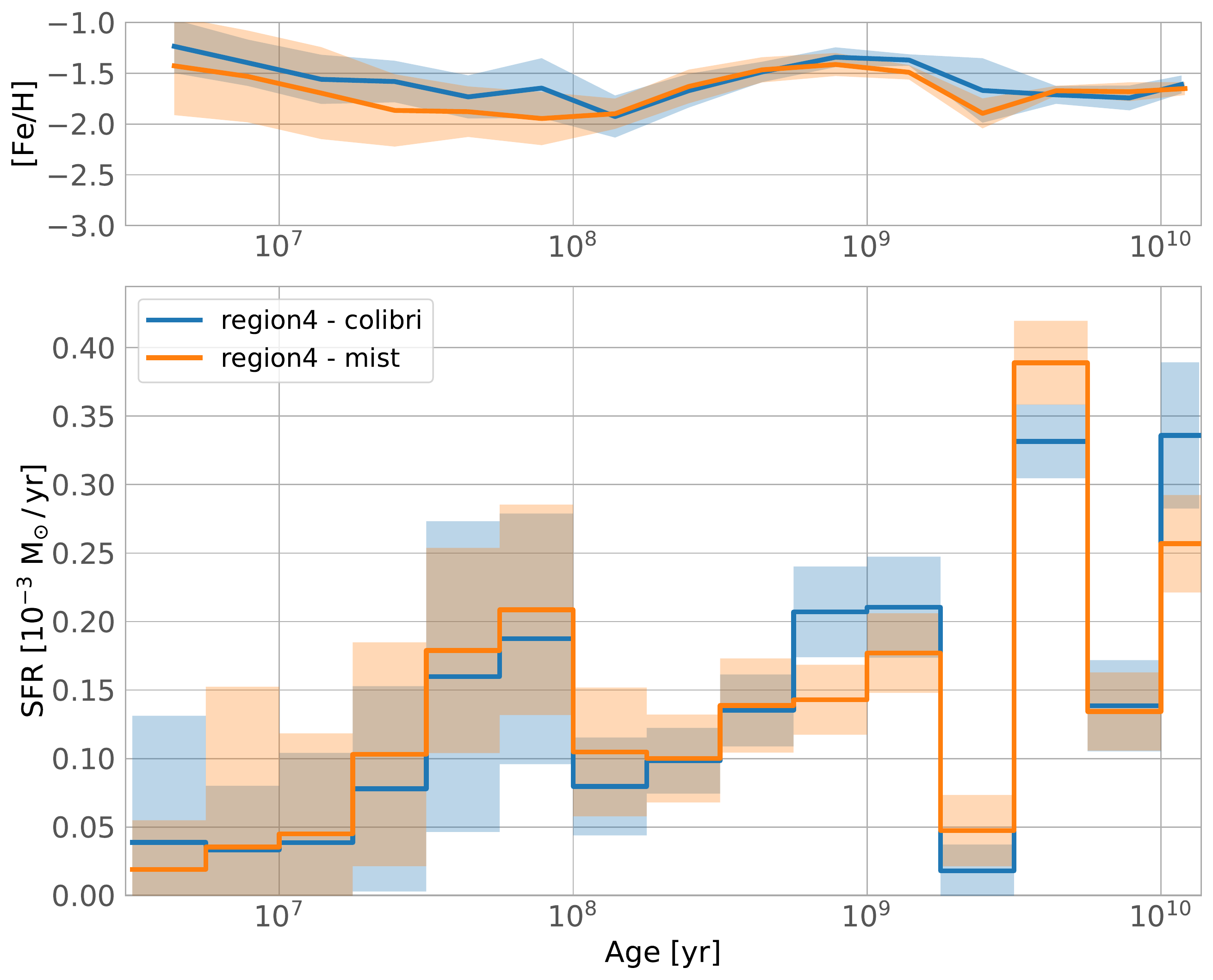}

\caption{\textit{Left panels.} Hess diagrams of the CMDs for the different regions (1 to 4 from top row to bottom row) of UGC~4483: the observational diagram is on the left and the one reconstructed on the basis of the MIST models in the middle, while on the right we show the residuals between the two in terms of the likelihood used to compare data and models with SFERA; the black line shows the 50\% completeness limit, used as a boundary for the SFH recovery. \textit{Right panels.} Recovered SFH from the two adopted sets of models (COLIBRI in blue, MIST in orange) with 1$\sigma$ error bars which include both random and systematic uncertainties.}
\label{fig:sfhs}
\end{figure*}

Going outwards from Region 1 to Region 4, we see a clear evolution of the CMD which becomes progressively older: the MS is less and less populated, while the RGB and RC features start to stand out. In Region 4 we reach a 50\% completeness at $\mathrm{m_{F814W}} \sim 28$. These differences will be reflected in the radial SFH of the galaxy.

\subsection{Region 1}
Region 1 includes the area just outside the H~\textsc{ii} regions of UGC~4483. Its CMD contains a variety of stellar populations, suggesting a continuous SF from ancient to recent epochs. Figure \ref{fig:sfhs}, top row, shows the Hess diagram (i.e. the density of points) of the CMD, with the observational diagram being on the left, the one reconstructed on the basis of the MIST models in the middle, and the residuals between the two on the right; the black line shows an estimate of the 50\% completeness limit, used as a boundary for the SFH recovery. The rightmost panel contains the best-fit SFH from the two adopted sets of models (blue for COLIBRI, orange for MIST).

As expected from the CMD analysis, we see a prevalence of young SF, even though the low number of stars and severe incompleteness result in huge errors on our rates. The models perfectly reproduce the good separation between the MS and blue BL sequences, and between the red BL and RGB sequences, a great value of these excellent data, and a significant improvement with respect to previous observations.

The two sets of models reasonably agree, in particular in the best constrained intermediate-age bins, between $\sim 60$ Myr and $\sim 1$ Gyr. In both cases, the best-fit SFHs were obtained with an additional internal reddening of $\mathrm{E(B-V)}=0.075$, i.e. $A_{\mathrm{F606W}} = 0.22$ and $A_{\mathrm{F814W}} = 0.14$.

\subsection{Region 2}
The results for Region 2 are shown in the second row of Figure \ref{fig:sfhs}. From the Hess diagram we can see a clear shortening of the bright MS and a higher density of older BLs, while the RGB starts to become more populated and more carbon  stars and TP-AGB stars appear. This is well reflected in the SFH, whose peak is between $\sim 30$ and $\sim 60$ Myr ago, with rates that stay high until $\sim 1$ Gyr ago, in both solutions. We still see SF in older times bins, but 
the increasing incompleteness and the worse time resolution make our results at the older epochs less reliable.

The internal reddening we find in this region is $\mathrm{E(B-V)}=0.075$, i.e. $A_{\mathrm{F606W}} = 0.22$ and $A_{\mathrm{F814W}} = 0.14$.

\subsection{Region 3} \label{sec:r3}
The decreasing young-to-old SFR trend continues in Region 3, as shown in Figure \ref{fig:sfhs}, third row. We find a roughly constant SF, 
with some ups and downs in particular in the COLIBRI solution. The internal reddening for this region is $\mathrm{E(B-V)}=0.025$, i.e. $A_{\mathrm{F606W}} = 0.07$ and $A_{\mathrm{F814W}} = 0.05$, which is lower that in the Regions 1 and 2, where the higher recent SF activity creates more dust, thus we can expect a more severe extinction.

Interestingly, the two solutions show a significant discrepancy in the last time bin, with the COLIBRI models providing a rate about 3 times higher than the MIST models. This is most likely a systematic difference in the stellar tracks, and in the way that the most uncertain stellar evolution parameters (like overshooting) are treated. It is worth noticing that the COLIBRI tracks also produce a dip in the SFH between $\sim 2$ and $\sim 3$ Gyr ago that is much less pronounced in the MIST models, maybe balancing that old peak at ages older than 10 Gyr ago. This is why it is important to use different sets of models, to compare their systematics and provide feedback to keep improving our knowledge on stellar evolution.

We also stress that this age range is constrained by stars at the edge of the 50\%  completeness, and should be interpreted with caution.

\subsection{Region 4}

Region 4 includes the most external part of UGC~4483, and our most complete data set. The SFH here, shown in the bottom row of Figure \ref{fig:sfhs}, is compatible with 0 back to $\sim 30$ Myr ago, and its peak is at ages older than $\sim 3$ Gyr ago, according to both solutions. We find an internal reddening of $\mathrm{E(B-V)}=0.05$, i.e. $A_{\mathrm{F606W}} = 0.14$ and $A_{\mathrm{F814W}} = 0.09$. This is lower than in Regions 1 and 2 but counter-intuitively a bit higher than in Region 3; however, the sensitivity of the CMD-fitting procedure to reddening can be degenerate with distance, so these values should be taken as an indication. Moreover, this reddening difference corresponds to a magnitude difference of about 0.05 mag, which gives a distance difference of about 10 pc, consistent with a distance difference along the line-of-sight between the body and the halo of the galaxy. The very deep CMD reaches 50\% of completeness at $\mathrm{m_{F814W}} \sim 28$, thus fainter than where we expect to see the HB of the galaxy (see the isochrones in Figure \ref{fig:cmd_iso}). Despite the systematic differences we just discussed, both sets of models show a consistent SFR at all ages, including those older than 10 Gyr ago, putting an important constraint on the age of this galaxy outside the Local Group.

\begin{figure}
\centering
\includegraphics[width=\linewidth]{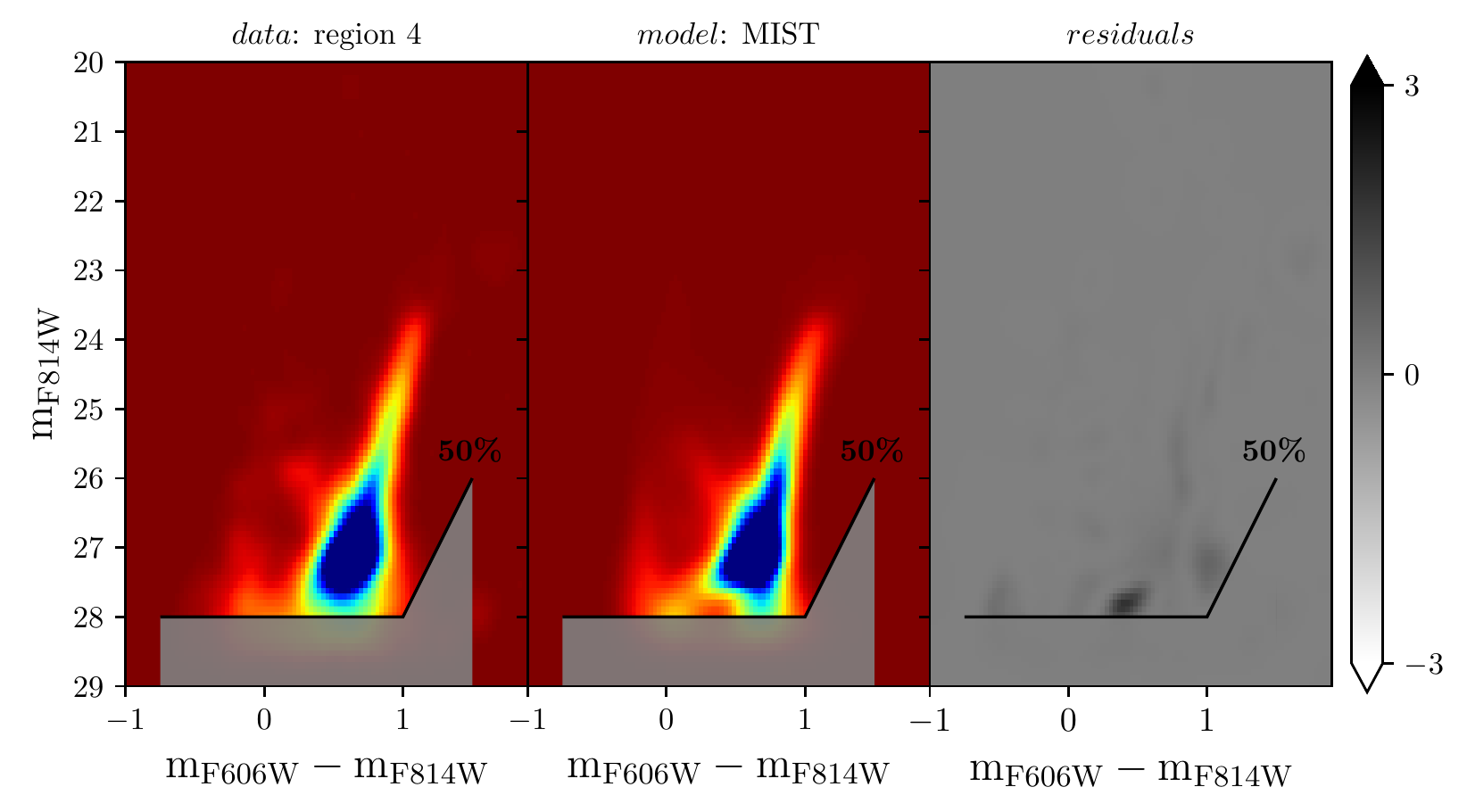}
\caption{Best-fit solution for Region 4, but with the SFH forced to stop 8 Gyr ago (to be compared with Figure \ref{fig:sfhs}, bottom row).}
\label{fig:r4_8Gyr}
\end{figure}

Since we are working at the edge of the data completeness, we tested the SFH recovery forcing the SF to start only 8 Gyr ago, instead of 13.7, to compare the results and check if such a SFH would still be compatible with the observed CMD. The result of this test is shown in Figure \ref{fig:r4_8Gyr}. Even though the difference with Figure \ref{fig:sfhs} might not seem significant, the recovered CMD, residuals, and output likelihood of this solution are significantly worse than the case shown in Figure \ref{fig:sfhs}, bottom row. Not only the faintest part of the RC/HB is not well reproduced, but also the RGB color does not match. This test provides a strong evidence that we indeed need a SFH older than 8 Gyr to fit our data; given the galaxy's low metallicity, we strongly believe that this points to the presence of an old HB, possibly leading to the existence of RRL variable stars in UGC~4483 (see Section \ref{sec:rrls} for a more detailed discussion).

If we shorten even more the lookback time at which we start the SF (e.g., 5 Gyr ago), the situation keeps worsening, as we remove from the synthetic CMD not only stars associated with the RC/HB, but also stars with younger ages that are present in the observed CMD. \\

It is worth mentioning that in all regions the best-fit solution is obtained by using a distance modulus DM~=~$27.45\pm0.10$, which is slightly lower than the one provided by \citet{Izotov2002}, but consistent with the one provided by \citet{Dolphin2001}, i.e. $27.53 \pm 0.12$.

A summary of the derived SFRs and stellar masses formed at various epochs in the different regions of the galaxy is given in Table \ref{tab:table}. In particular for Regions 3 and 4, we can notice a significant difference in the epoch of the main SF peak derived with the COLIBRI or the MIST models. As already discussed in Section \ref{sec:r3}, the 50\% completeness in Region 3 is at the edge of the HB or the faintest part of the RC phases, thus systematic uncertainties in the stellar evolution models can greatly affect the resulting age and rate determination, depending on how much of the synthetic HB falls within or outside the CMD region covered by the photometry. For Region 4, the situation is better, thanks to the lower crowding; despite the systematics still affecting the fit, Figure \ref{fig:sfhs} (bottom right panel) shows a very good agreement between the two solutions, and the difference in $\mathrm{age_{\, peak}}$ reported in Table \ref{tab:table} ($3-6$~Gyr for the MIST models, versus $10-13.7$~Gyr for the COLIBRI models) is simply the result of small variations of the SFRs in the last 3 age bins.

\subsection{Searching for RRL variable stars in UGC 4483}\label{sec:rrls}
As already mentioned, the detection of RRL stars in a galaxy can unambiguously confirm the presence of an HB, and unveil a stellar population at least $\sim 10$~Gyr old. RRLs are $\sim 3$ mag brighter than coeval turnoff stars, and the typical form of their light variation makes them easily recognized even in very crowded fields and in galaxies where a 10 Gyr population may be buried into the younger stars. To look for these variables in UGC~4483, the observations have been reduced as a time series to provide a full mapping of the classical instability strip of pulsating variables with periods from about half a day to a couple of days. A detailed analysis of the candidate RRLs and theirs light curves will be the subject of a forthcoming paper (Garofalo et al., in preparation), while here we simply mention what is relevant to the present paper.

\begin{figure}
\centering
\includegraphics[width=\linewidth]{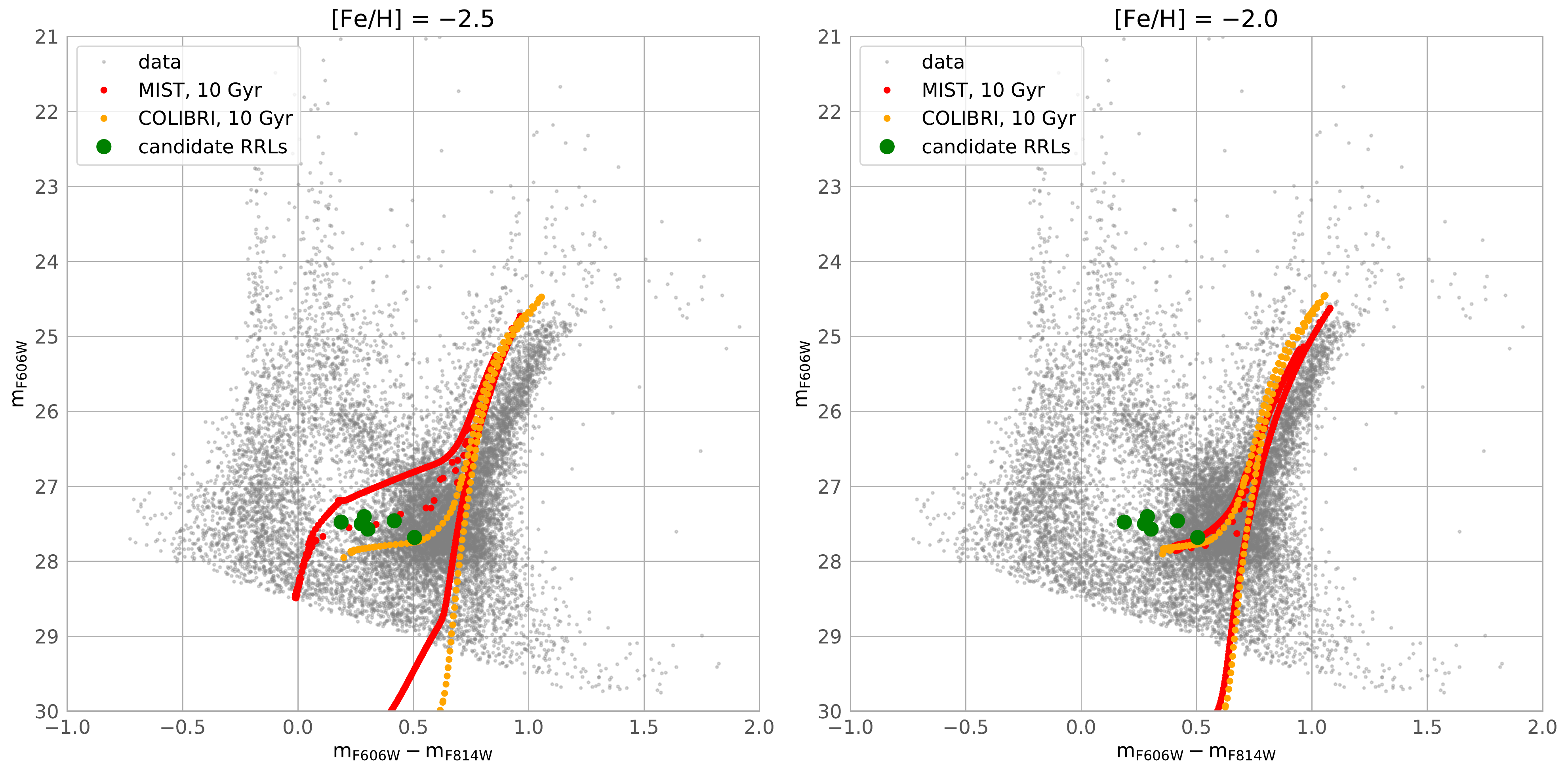}
\caption{F606W versus F606W$-$F814W color-magnitude diagram of UGC~4483 in grey, the six RRL candidates we found in green, and the MIST (in red) and COLIBRI (in orange) 10 Gyr old isochrones for a metallicity of [Fe/H] = $-2.5$ (left panel) and [Fe/H] = $-2.0$ (right panel). The isochrones were shifted to match the distance and foreground extinction of the galaxy.}
\label{fig:rrls}
\end{figure}

In our analysis, we found six stars with light curves compatible with those of RRL variables. In particular, two of them are in the most external region of the galaxy we analyzed, Region 4, where crowding conditions are less severe and the completeness much more favorable than in more internal regions. Within the uncertainties of the HB modeling, these candidates also have colors and magnitudes compatible with being indeed RRLs, as shown in Figure \ref{fig:rrls}. In the left panel, we plot the F606W versus F606W$-$F814W color-magnitude diagram of UGC~4483 in grey, the six candidates in green, and the MIST (in red) and COLIBRI (in orange) 10 Gyr old isochrones for a metallicity of [Fe/H] = $-2.5$. We can notice how different the models are, with the MIST HB being more than half a magnitude brighter than the COLIBRI one. At this metallicity, our candidates fall between the two sets of HB. As shown in the right panel, the disagreement between the isochrones starts to disappear at [Fe/H] = $-2.0$, and most of our candidates would fall slightly brighter and bluer than the HB, but still compatible with it once the uncertainties on both data and models are considered.


\begin{figure}
\centering
\includegraphics[width=\linewidth]{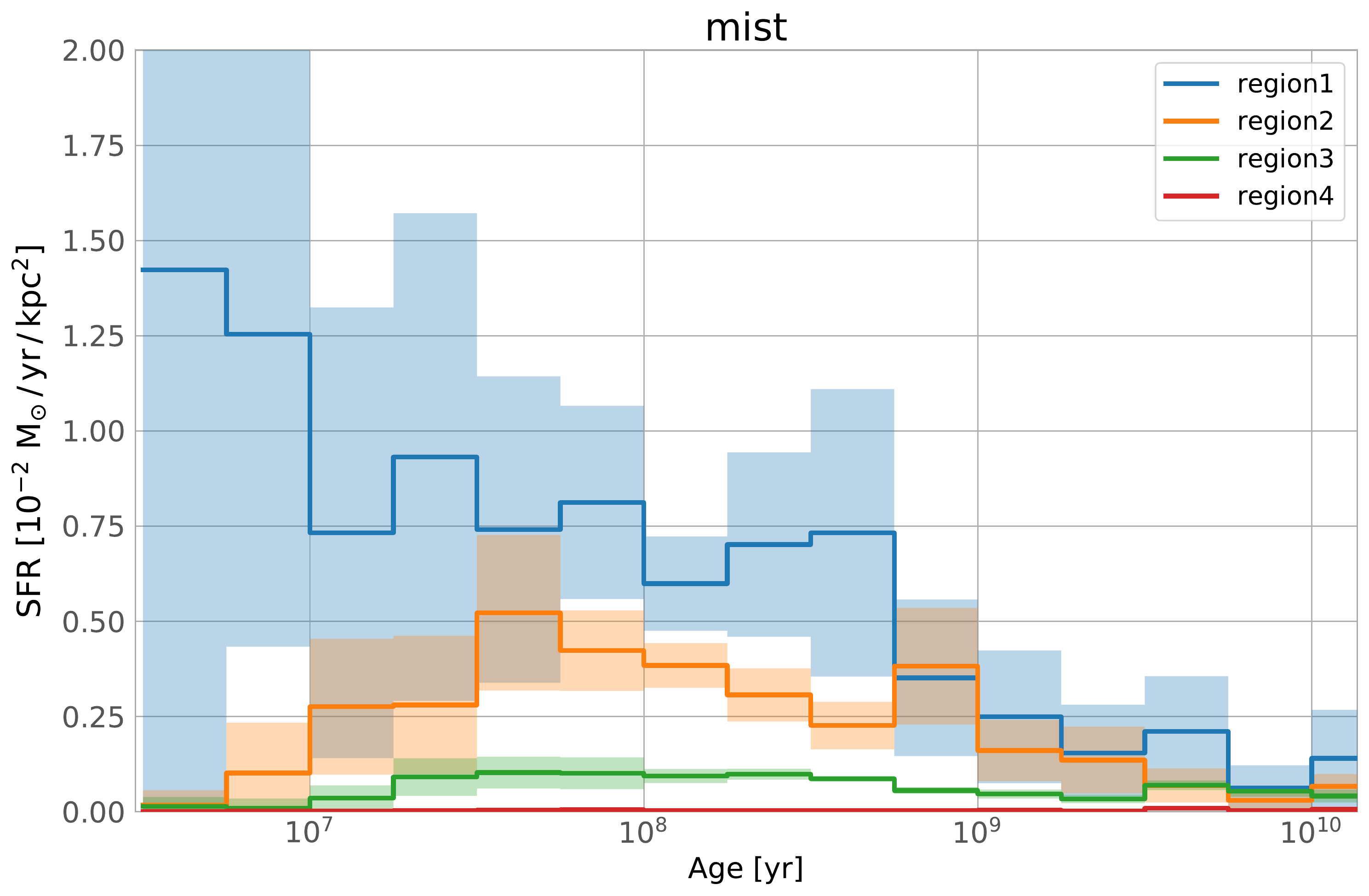}
\caption{SFR surface densities (SFR/area) as a function of time in different regions of UGC~4483, from the MIST solutions.}
\label{fig:cfr_area}
\end{figure}


\begin{table*}[ht]
\caption{Summary of the derived star formation rates and stellar masses in the different regions of UGC~4483.\\
$^{(*)}$ The last column is the ratio between the two previous ones.}
\begin{center}
\begin{tabular}{ccccccc}
    \toprule
    \midrule
    \addlinespace[0.3em]
    \multirow{2}{*}{region} & $\mathrm{\langle SFR \rangle}$ & $\mathrm{SFR_{\, peak}}$ & \multirow{2}{*}{$\mathrm{age_{\, peak}}$} & $\mathrm{M_{\ast}(age \leq 50\ Myr)}$ & $\mathrm{M_{\ast}(age > 1\ Gyr)}$ & young/old SFR $^{(*)}$\\
    \addlinespace[0.3em]
    & $\mathrm{[M_{\odot}/yr/kpc^2]}$ & $\mathrm{[M_{\odot}/yr/kpc^2]}$ & & $\mathrm{[10^4\ M_{\odot}]}$ & $\mathrm{[10^6\ M_{\odot}]}$ & [$10^{-2}$]\\
    \midrule
     \multicolumn{7}{c}{\textbf{COLIBRI models}}\\
    \midrule
    1 & $(1.73 \pm 0.64) \times 10^{-3}$ & $(2.10 \pm 1.83)  \times 10^{-2}$ & $0-6$ Myr & $2.34 \pm 0.79$ & $1.25 \pm 0.55$ & 1.88 \\
    2 & $(1.11 \pm 0.26) \times 10^{-3}$ & $(4.98 \pm 0.20) \times 10^{-3}$ & $30-60$ Myr & $3.13 \pm 0.92$ & $1.97 \pm 0.56$ & 1.59 \\
    3 & $(6.87 \pm 0.96) \times 10^{-4}$ & $(1.30 \pm 0.28) \times 10^{-3}$ & $10-13.7$ Gyr & $1.53 \pm 0.45$ & $3.38 \pm 0.52$ & 0.45 \\
    4 & $(4.83 \pm 0.40) \times 10^{-5}$ & $(7.38 \pm 1.17) \times 10^{-5}$ & $10-13.7$ Gyr & $0.56 \pm 0.30$ & $2.85 \pm 0.25$ & 0.19 \\
    \midrule
     \multicolumn{7}{c}{\textbf{MIST models}}\\
    \midrule
    1 & $(1.66 \pm 0.54) \times 10^{-3}$ & $(1.42 \pm 1.46)  \times 10^{-2}$ & $0-6$ Myr & $2.88 \pm 0.94$ & $1.08 \pm 0.45$ & 2.68 \\
    2 & $(8.66 \pm 1.92) \times 10^{-4}$ & $(5.22 \pm 2.04) \times 10^{-3}$ & $30-60$ Myr & $3.08 \pm 0.93$ & $1.36 \pm 0.40$ & 2.27 \\
    3 & $(5.19 \pm 0.74) \times 10^{-4}$ & $(1.02 \pm 0.42) \times 10^{-3}$ & $30-60$ Myr & $1.62 \pm 0.50$ & $2.51 \pm 0.40$ & 0.65 \\
    4 & $(4.54 \pm 0.34) \times 10^{-5}$ & $(8.54 \pm 0.67) \times 10^{-5}$ & $3-6$ Gyr & $0.64 \pm 0.23$ & $2.70 \pm 0.21$ & 0.24 \\
    \midrule
    \bottomrule
\end{tabular}
\end{center}
\label{tab:table}
\end{table*}

\section{Discussion}
The Local Group is an incredible environment, full of interesting processes and galaxies of many different types. However, since there are not known BDCs within it (except for IC~10, sometimes considered as a such), in order to study this class of very actively star-forming, low-metallicity, systems the only possibility is to push our limits beyond its borders. Even though the distance makes observations challenging even for \textit{HST}, with the photometry reaching only the brightest evolutionary phases, many studies tried to characterize the formation and evolution of these intriguing objects.

Among these works, some embarked on the challenge of applying the synthetic CMD method to the observed CMDs, trying to reconstruct the galaxies' SFHs within the reachable lookback time (see, e.g., \citealt{Aloisi1999} for I~Zw~18, \citealt{Schulte-Ladbeck2001} for I~Zw~36, or \citealt{Annibali2003} for NGC~1705). Despite the obvious differences among the individual SFHs, all these dwarfs show a qualitatively similar behavior, with a strong ongoing SF activity and a moderate and rather continuous star formation at older epochs. Most notably, they all show an RGB, meaning that they contain stars as old as the lookbacktime reached by the photometry. Finding old stars in these extremely metal-poor galaxies is a key information to understand how they compare to other systems and to place them in the general context of galaxy formation and evolution. For this reason, it is crucial to try to reach even older stellar evolutionary features, to finally figure out whether BCDs are really old systems as currently believed.

We presented here a detailed analysis of UGC~4483, the closest example of the BCD category, which we studied thanks to deep new data obtained with the exact purpose of reaching stars older than a few Gyr. Our new WFC3/UVIS observations reach more than 4 mag deeper than the TRGB, and allow us to detect and characterize for the first time the population of core He-burning stars with masses $\lesssim 2$~M$_{\odot}$.

To take into account the different crowding and SF conditions across the galaxy, we used isophotal contours of the F606W image to divide it into sub-regions, as shown in Figure \ref{fig:regions}; the resulting SFHs from the MIST solution are shown in Figure \ref{fig:cfr_area}, where the rates have been normalized to each region's area to facilitate the comparison. As expected for a BCD, the main recent activity is in the central Region 1, which, despite the uncertainties, shows a continuous increase of SF in the most recent time bins, with a rate about 10 times higher than the average. Going outwards, the rate densities generally decrease, together with the ratio of present-to-past average SF (see the last column of Table \ref{tab:table} for the details). Indeed, Regions 3 and 4 had already formed 50\% of their total stellar mass around $\sim 5$~Gyr ago, while this occurred between 2 and 3~Gyr ago for Regions 1 and 2, confirming that the SF proceeded from outside in.

Even if we could not derive the SFHs there, it is clear that this trend still holds when we consider the two innermost regions of the galaxy, 0a and 0b, sites of very bright and active H\textsc{ii} regions. In particular, Region~0a hosts a super star cluster with absolute magnitudes of $M_V = -8.98 \pm 0.24$ and $M_I = -9.06 \pm 0.16$, corresponding to an age of $\sim 15$~Myr and an initial mass of $\sim 10^4$~M$_{\odot}$, as derived by \citet{Dolphin2001}. For comparison, this is just slightly fainter than 30 Doradus, the star forming region within the Large Magellanic Cloud. Using the equivalent width of the H$\beta$ emission line, \citet{Izotov2002} derived a much younger age of 4 Myr.

In the outer part of the galaxy, Region 4, we find the SF peak at ages $\gtrsim 3$~Gyr, and the rates are consistent with a SFH starting 13.7 Gyr ago. This is the first time that we can put such a strong constraint on a galaxy outside the Local Group, and despite the uncertainties and systematics of the stellar evolution models, we believe this to be a robust result which is supported also by the detection of a number of candidate RR Lyrae stars. The implication is that UGC~4483, and possibly all BCD galaxies, are indeed as old as a Hubble time, and not young systems as previously believed. 
Indeed, despite the very young recent activity, about 87\% of the total stellar mass of the galaxy was formed at ages older than 1~Gyr.

The color extension of the identified HB region includes the expected location of the RR Lyrae gap, corresponding to the pulsation instability strip. Follow-up time-series observations would be very useful to definitely confirm our detection in UGC~4483 (Garofalo et al., in preparation) of candidate pulsating stars of this class (unambiguous tracers of a $\gtrsim 10$ Gyr old population) and to be able to use the inferred pulsation characteristics (periods, amplitudes, etc.) to trace the properties of the oldest stellar population in UGC~4483.

It is interesting to notice that, as far as the SFH is concerned, UGC~4483 is fairly similar to many dIrrs, in spite of being classified a BCD: it shows a rather continuous SF, with no evidence of short, intense bursts, and no long quiescent phases, at least within the limits of our time resolution. This adds to the evidence that on average the SFH is similar in the two classes of star-forming dwarfs, and the somewhat old idea that BCDs are experiencing now much higher SF activities than dIrrs is not supported by their CMDs. We can indeed compare UGC~4483 to the cases of, e.g., NGC~1569 \citep{Grocholski2012} or NGC~4449 \citep{Sacchi2018}, both very active dIrrs.

The extremely low metallicity of UGC~4483 makes our result of the SF being active throughout the whole Hubble time even more important in terms of galaxy chemical evolution. Indeed, this system has been making stars, therefore synthesizing heavy elements for all its long life, and yet it is still extremely metal poor today. This implies not only that significant infall of metal poor gas must have occurred, but also that other mechanisms, such as outflows/winds removing heavy elements, are necessary.

It is interesting to check how our results compare to the literature, in particular with previously derived SFHs. Based on \textit{HST}/WFPC2, \citet{Dolphin2001} derived a SFR of $1.3 \times 10^{-3}~\mathrm{M_{\odot}/yr}$, which is almost a factor of two higher than ours, $(7.01\pm0.44) \times 10^{-4}$~$\mathrm{M_{\odot}/yr}$. However, their CMD is not particularly deep (the faintest stars have an I magnitude of 25, with much larger photometric errors), so it is likely that they are biased by the brightest young stars in the upper part of the CMD. On the other hand, our rate is a lower limit as it is based on Regions 1 to 4, thus excluding the most active Regions 0a and 0b which have not enough stars to run a full SFH derivation. Based on the same data, \citet{McQuinn2010} derived a more detailed SFH, which shows a very recent SF peak (within the last 50 Myr), consistent with what we find in Region 1, and a rather continuous activity up to 1 Gyr ago. Their peak rate, however, is about  $1.1 \times 10^{-2}~\mathrm{M_{\odot}/yr}$, while we reach at most $2.5 \times 10^{-3}~\mathrm{M_{\odot}/yr}$ when considering the uncertainties. This can be again the result of the different sampling of the stellar populations, in particular in the most central regions of the galaxy. They also find an old population, which is consistent with the RGB phase revealed by their CMD, and confirmed by our deeper one. Also \citet{Weisz2011} used the same data and code to derive the SFH of this galaxy, obtaining similar results (their average SFR of $1.57 \times 10^{-3}~\mathrm{M_{\odot}/yr}$).

From the H$\alpha$ luminosity \citep{GildePaz2003,Kennicutt2008}, we can infer the current ($\lesssim 10$~Myr) SFR following the prescription of \citet{Murphy2011}, which assumes a Kroupa IMF. We obtain $2.2 \times 10^{-3}~\mathrm{M_{\odot}/yr}$, compatible within the error bars with our results in the most recent time bin of Region 1, but still higher, as expected as the H$\alpha$ mainly traces the emission from the two most active Regions 0a and 0b (see Figure \ref{fig:image}, right panel).

Finally, from the analysis of VLA observations, \citet{Lelli2012b} found that UGC 4483 has a steeply-rising rotation curve, making it the lowest-mass galaxy with a differentially rotating H\textsc{i} disk. These rotation-velocity gradients are directly related to the dynamical mass surface densities, and signal a strong central concentration of mass, also found in other BCDs, like I Zw 18 \citep{Lelli2012a}, NGC 2537 \citep{Matthews2008}, and NGC 1705 (\citealt{Meurer1998}; see also Figure 10 of \citealt{Lelli2012b}). They also showed that the central mass concentration cannot be explained by the newly formed stars or by the concentration of the H\textsc{i}, implying that either the progenitors of BCDs are compact, gas-rich dwarfs, or there must be a mechanism (external, such as interactions and mergers, or internal, such as torques from massive star-forming clumps, \citealt{Elmegreen2012}) leading to a concentration of gas, old stars, and/or dark matter, causing the SF increase. 



\section{Summary and Conclusions}
Here we summarize the main results of this paper, where we presented new WFC3/UVIS data of the BCD galaxy UGC 4483, and investigated its resolved stellar populations and radial SFH.

\begin{itemize}
    \item[-] The CMD of UGC~4483 is populated by many generations of stars, from young MS and BL stars, to intermediate-age AGB and RGB stars, and older RC and possibly HB stars. In particular, we were able to reach and detect for the first time the population of core He-burning stars with masses $\lesssim 2$~M$_{\odot}$.
    \item[-] The stellar populations in the galaxy have the typical distribution of star-forming irregular galaxies, with the youngest stars being more centrally concentrated and close to the inner star-forming regions, while the distribution becomes more and more uniform as the stars age.
    \item[-] From the SFH recovered in different radial regions of the galaxy, we found a declining trend of the overall SF activity and of the present-to-past SFR ratio going from inside out. In all regions, we found the best fit SFH using a distance modulus of DM~=~$27.45\pm0.10$, slightly lower than previous estimates.
    \item[-] Using the synthetic CMD method, we determined an average SFR over the whole Hubble time of $(7.01\pm0.44) \times 10^{-4}$~$\mathrm{M_{\odot}/yr}$, corresponding to a total astrated stellar mass of $(9.60\pm0.61)\times 10^6$~$\mathrm{M_{\odot}}$, 87\% of which went into stars at epochs earlier than 1~Gyr ago. These are lower limits, as they do not include the two innermost regions of the galaxy, hosting very active H~\textsc{ii} regions and a super star cluster almost as luminous as 30 Doradus.
    \item[-] We found strong evidence of a $\gtrsim 10$~Gyr old population, which might be responsible for the presence of a blue HB, as also suggested by our detection of a number of candidate RRL variable stars in UGC~4483.
\end{itemize}

\acknowledgments{
These data are associated with the \textit{HST} GO Program 15194 (PI A. Aloisi). Support for this program was provided by NASA through grants from the Space Telescope Science Institute. F.A., M.C., and M.T. acknowledge funding from the INAF Main Stream program SSH 1.05.01.86.28. We thank the anonymous referee for the useful comments and suggestions that helped to improve the paper.
}

\bibliography{bib}

\end{document}